\documentclass[preprint2]{aastex}

\begin{document}

\slugcomment{To appear in the Astronomical Journal (Sept 2005)}
\shortauthors{Fitzpatrick \& Massa}
\shorttitle{Extinction Without Standards}
\newcommand{\atlas}{{\it ATLAS9}}
\newcommand{\synspec}{{\it SYNSPEC}}
\newcommand{\iras}{{\it IRAS}}
\newcommand{\hst}{{\it HST}}
\newcommand{\iue}{{\it IUE}}
\newcommand{\oao}{{\it OAO-2}}
\newcommand{\td}{{\it TD-1}}
\newcommand{\tmass}{{\it 2MASS}}
\newcommand{\hip}{{\it Hipparcos}}
\newcommand{\vsini}{$v \sin i$}
\newcommand{\kms}{km\,s$^{-1}$}
\newcommand{\teff}{{$T_{\rm eff}$}}
\newcommand{\logg}{{$\log g$}}
\newcommand{\vturb}{$v_{turb}$}
\newcommand{\ebv}{$E$(\bv)}
\newcommand{\msun}{${\rm M}_\sun$}

\title{An Analysis of the Shapes of Ultraviolet Extinction Curves. IV.
Extinction without Standards\\ {\small(To appear in the September 2005 Astronomical Journal)}}

\author{E.L.~Fitzpatrick\altaffilmark{1}, D.~Massa\altaffilmark{2}}
\altaffiltext{1}{Department of Astronomy \& Astrophysics, Villanova
University, 800 Lancaster Avenue, Villanova, PA 19085, USA; fitz@astronomy.villanova.edu}
\altaffiltext{2}{SGT, Inc., NASA/GSFC, Mailstop 681.0, Greenbelt,
MD 20771; massa@derckmassa.net}

\begin{abstract}
In this paper we present a new method for deriving UV-through-IR
extinction curves, based on the use of stellar atmosphere models to
provide estimates of the intrinsic (i.e., unreddened) stellar spectral
energy distributions (SEDs), rather than unreddened (or lightly
reddened) standard stars.  We show that this
``extinction-without-standards'' technique greatly increases the
accuracy of the derived extinction curves and allows realistic
estimations of the uncertainties.  An additional benefit of the
technique is that it simultaneously determines the fundamental
properties of the reddened stars themselves, making the procedure
valuable for both stellar and interstellar studies.  Given the physical
limitations of the models we currently employ, the technique is limited
to main sequence and mildly-evolved B stars.  However, in principle, it
can be adapted to any class of star for which accurate model SEDs are
available and for which the signatures of interstellar reddening can be
distinguished from those of the stellar parameters. We demonstrate how
the extinction-without-standards curves make it possible to: 1) study
the uniformity of curves in localized spatial regions with
unprecedented precision;  2) determine the relationships between
different aspects of curve morphology; 3) produce high quality
extinction curves from low color excess sightlines; and 4) derive
reliable extinction curves for mid-late B stars, thereby increasing
spatial coverage and allowing the study of extinction in open clusters
and
associations dominated by such stars.  The application of this
technique to the available database of UV-through-IR SEDs, and to
future observations, will provide valuable constraints on the nature of
interstellar grains and on the processes which modify them, and will
enhance our ability to remove the multi-wavelength effects of
extinction from astronomical energy distributions.
\end{abstract}

\keywords{ISM:dust,extinction --- stars:atmospheres,abundances ---
methods:data analysis}


\section{INTRODUCTION}
A detailed determination of the wavelength dependence of interstellar
extinction, i.e., the absorption and scattering of light by dust grains,
is important for two very different reasons.  First, since it is a
product of the optical properties of dust grains, extinction provides
critical diagnostic information about interstellar grain populations
(including size distribution, grain structure, and composition),
providing guidance for interstellar grain models.  Second, the accuracy
to which the intrinsic spectral energy distributions (SEDs) of most
astronomical objects can be determined depends on how well the effects
of extinction can be removed from observations.  In both cases, a
fundamental issue is how accurately the wavelength dependence of
extinction can be measured.  Consequently, it is essential to have a
firm grasp of how measurement errors can affect the determination of
this wavelength dependence.

This paper is the culmination of a series of ``techniques'' papers
published over the past five years (Fitzpatrick \& Massa 1999; Massa \&
Fitzpatrick 2000; and Fitzpatrick \& Massa 2005; hereafter FM99, MF00,
and FM05, respectively) whose aim has been to develop a technique to
{\it simultaneously determine} the wavelength dependence of extinction
(to higher accuracy than previously possible) and the physical
properties of a reddened star.  It represents a continuation of our
earlier series on the properties of UV extinction (Fitzpatrick \& Massa
1986; 1988; and 1990, hereafter FM90), and provides a detailed
description of a new technique for deriving interstellar extinction
curves which does not rely on observations of standard stars, virtually
eliminates the effects of ``mismatch'' error, and yields an accurate
assessment of the uncertainties.  This ``extinction-without-standards''
technique opens the door to a new class of extinction studies,
including regions heretofore inaccessible.  For example, errors in the
traditional ``pair method'' approach to extinction strongly limit our
ability to study extinction in two important regimes. The first is
low-$E(B-V)$ sightlines, where one might hope to relate extinction
properties to the environmental properties of specific physical
regions.  The second is extinction derived from mid- to late B stars.
These stars are plentiful, and often constitute the bulk of the stars
available for extinction measurements in intrinsically interesting
regions, such as the Pleiades.  We will demonstrate how the
extinction-without-standards approach overcomes both of these problems
and present examples of each.  Some first results from this program
were illustrated by Fitzpatrick (2004, hereafter F04).
In addition to the extinction results, we will demonstrate that our
approach simultaneously provides accurate stellar parameters which are
also astrophysically interesting.

In \S\ref{measure}, we provide a broad overview of the basic problem of
measuring an extinction curve, and compare the merits of curves derived
by the pair method and curves derived by using stellar atmosphere
models.  In \S\ref{secEXTINCTION}, we describe our new model-based
technique in detail, and list the basic ingredients needed to determine
an extinction curve using this approach.  In \S\ref{secDATA}, we
describe the data used in the current study.  In \S\ref{secRESULTS} we
provide a number of sample extinction curves derived using model
atmospheres.  We also demonstrate the high precision of the new curves
and the reliability of the error analysis employed.  Finally in
\S\ref{secDISCUSS} we describe some to the scientific advantages of
this new technique and our plans to exploit them.


\section{MEASURING EXTINCTION\label{measure}}
To understand how interstellar extinction is measured and to appreciate
how different measurement techniques can affect the outcome, we begin
with the intrinsic elements of an uncalibrated observation of a spectral
energy distribution (SED) of a reddened star obtained at the earth,
$f_{\lambda}$.  This can be expressed as
\begin{equation}
f_{\lambda} = F_{\lambda}\, r_\lambda\, \theta_R^2\, 10^{-0.4
A_{\lambda}}
\label{flux1}
\end{equation}
where $F_{\lambda}$ is the intrinsic surface flux of the star at
wavelength $\lambda$, $r_\lambda$ is the response function of the
instrument, $\theta_R \equiv (R/d)^2$ is the angular radius of the star
(where $d$ is the stellar distance and $R$ is the stellar radius), and
$A_{\lambda}$ is the absolute attenuation of the stellar flux by
intervening dust (i.e., the total extinction) at $\lambda$.
Alternatively, the observed SED can be expressed in terms of magnitudes
$m_{\lambda}$ by
\begin{eqnarray}
m_{\lambda} &= &-2.5 \log F_{\lambda} \theta_R^2  +A_{\lambda}
+C_\lambda \;\; \;\; \;\; {\rm or}
\label{mag1} \\
m_{\lambda} &= &M_{\lambda} +5 \log d -5 +A_{\lambda} +C_\lambda
\label{mag1b}
\end{eqnarray}
where $C_\lambda = -2.5 \log r_\lambda$ is a term which transforms
between the observed magnitude system and absolute flux units, and
$M_\lambda$ is the traditional definition of the absolute magnitude of
the star at $\lambda$.

The difficulty in measuring the total extinction $A_{\lambda}$ can be
seen
by rearranging these equations to solve for the extinction term.
Equation~\ref{flux1} yields
\begin{equation}
A_{\lambda} = 2.5 \log (r_\lambda \theta_R^2
              \frac{F_{\lambda}}{f_{\lambda}}) \;\;\;\;\;\;  ,
\label{alam1}
\end{equation}
while Equation \ref{mag1b} yields
\begin{equation}
A_{\lambda} = m_\lambda -M_\lambda -5 \log d  +5 +C_\lambda
\;\;\;\;\;\;  .
\label{alam2}
\end{equation}
In either case, a true measurement of $A_\lambda$ would require
calibrated SED observations, knowledge of the intrinsic SED of the
star, and measurements of both the stellar distance and radius, or
their ratio $\theta_R$.  Unfortunately, there are no early-type stars
for which both of these latter quantities are known to sufficient
accuracy to allow a meaningful measurement of $A_{\lambda}$.  As a
result, indirect methods must be employed, and $A_\lambda$ is {\em
always a derived quantity}, subject to assumptions.

In virtually all extinction studies, the actual measured quantity is a
``color excess'' which describes the extinction at a wavelength
$\lambda$ relative to that at a fiducial wavelength.  The traditional
approach is to adopt the $V$ band as the fiducial, since $V$ magnitudes
are widely available, accurately calibrated, and typically of high
quality.  This color excess can be expressed as either
\begin{equation}
E(\lambda -V) \equiv A_{\lambda} - A_{V} = 2.5 \log
(\frac{r_\lambda}{r_V}
\frac{F_\lambda}{F_V} \frac{f_V}{f_\lambda}) \;\;\;\;\;\;  ,
\label{elam1}
\end{equation}
as based on Equation \ref{alam1}, or
{\small
\begin{eqnarray}
E(\lambda -V)  & \equiv & A_{\lambda} - A_{V} \nonumber \\
                                       & = & (m_\lambda -m_V) -
                                              (M_\lambda - M_V) +
                                              (C_\lambda - C_V)
					      \nonumber \\
                                        & = & m(\lambda-V) -
					M(\lambda-V) +
                                              C(\lambda-V)
					      \;\;\;\;\;\;  ,
\label{elam2}
\end{eqnarray}
}
as based on equation \ref{alam2}.  Thus, the determination of the color
excess requires only the measurement of the observed SED and a
knowledge of the shape of the intrinsic SED.  There are two basic
approaches to determining color excesses, based on the use of either
unreddened stars or stellar atmosphere models to represent the
intrinsic SEDs of reddened stars. These two techniques, and the issue
of the normalization of extinction curves, are discussed in the three
subsections to follow.

\subsection{The Pair Method}
The first approach is the ``pair method.''  A pair method curve is
constructed by comparing the fluxes of a reddened star and an (ideally)
identical unreddened ``standard star.'' Essentially, the
absolute magnitudes in Eq. \ref{elam2} are replaced by the observed
magnitudes of the standard star and the curve is usually expressed in
the
form
{\small
\begin{equation}
k(\lambda-V) \equiv \frac{E(\lambda-V)}{E(B-V)} \equiv
\frac{m(\lambda -V)-m(\lambda -V)_0}{(B-V) - (B-V)_0}   \;\;\;\;\;\;  ,
\label{klambda}
\end{equation}
}
where the color excesses $E(\lambda-V)$ are normalized by \ebv\/ and the
subscripted quantities refer to unreddened indices for the standard
star.  (The issue of normalization will be discussed below.)  When the
two stars are observed using the same instrument, this method has the
advantage that calibration terms cancel, eliminating any dependence on
the absolute flux calibration, $r_\lambda$ or $C_\lambda$.  There are,
however, two disadvantages to this technique.  The first is that the
grid of unreddened standard stars is necessarily limited, so that
some mismatch in the SEDs of the reddened and unreddened stars, termed
``mismatch error'', is inevitable.  The second is that there are very
few truly unreddened early-type stars, so usually the ``unreddened''
standard must be corrected for some small amount of extinction whose
exact magnitude and wavelength dependence are uncertain.  This creates
an error that can propagate into the resulting extinction curves.

Massa, Savage, \& Fitzpatrick (1983) presented a detailed study of the
uncertainties affecting pair method extinction curves and showed that
mismatch effects dominate the error budget.  If one uses a single
unreddened spectral standard for each spectral class, and assumes that
all spectral classifications are perfect, then as a result of spectral
binning, mismatch errors will be equal to or less than half a spectral
class.  Figure~\ref{figMISMATCH} shows how such mismatches can affect
extinction curves derived from main sequence B stars.  In each of the
four groups of curves in Figure~\ref{figMISMATCH}, the true shape of an
extinction curve affecting a B1, B2, B5, or B9 star is
indicated by the solid curve.  The extinction curves which would be
derived via the pair method in the presence of a $\pm\frac{1}{2}$
spectral class mismatch error are shown by the dash-dot curves for the
case of $E(B-V)=0.15$ and by the dashed curves for $E(B-V)=0.30$.
(The derived curves fall below the true curve in the UV and above the
true curve in the IR when the standard star is cooler than the reddened
star, and vice versa.)  In practice, spectral classifications are not
perfect and the available unreddened standard stars do not necessarily
lie in the middle of the range of properties within a single spectral
class.  Both these effects exacerbate the spectral mismatch problem and
thus the uncertainties shown in Figure ~\ref{figMISMATCH} are likely
closer to typical errors, rather than extremes.

Although it may appear from Figure \ref{figMISMATCH} that a
discontinuity at the Balmer jump at $\lambda^{-1} \simeq 2.7 \mu{\rm
m}^{-1}$ would provide an obvious indicator of the presence of spectral
mismatch in an extinction curve, this is almost never practical.  The
spectrophotometric data available for constructing extinction curves
are generally limited to UV wavelengths ($\lambda^{-1} > 3.3 \mu{\rm
m}^{-1}$), which are highlighted in Figure \ref{figMISMATCH}.
Typically, the only data available in the optical and near-UV are
photometric indices which straddle the Balmer jump, such as the Johnson
$U-B$ color, and these cannot be used to distinguish the effects of
mismatch from intrinsic curve shape.

Mismatch error clearly can have a profound effect on the shapes of
curves
derived from stars with low color excesses, particularly in the
mid-to-late B spectral range and most particularly in the UV spectral
region.  In fact, it is mismatch error which provides the low
temperature
cutoff to the spectral range of stars from which useful UV extinction
measurements can be made.  Mismatch also severely limits extinction
studies based on stars hotter and/or more luminous than the main
sequence B stars.  For the O stars, very few unreddened standard stars
exist and extreme mismatching of spectral types is often necessary to
derive a curve --- although, since the intrinsic UV/optical SEDs of the
O stars are not well known, it is not clear how large an effect this
introduces in the derived curves. The use of luminous, evolved stars
for
extinction studies is particularly problematic since unreddened
standards are rare and the sensitivity of the intrinsic SEDs to both
temperature and surface gravity can lead to very severe mismatch
effects in the resultant curves (although see the discussion in
Cardelli,
Sembach, \& Mathis 1992 for results in the early-B spectral range).

\subsection{Model Atmosphere Techniques}

The second approach to deriving extinction curves is to model the
intrinsic SED of the reddened star in order to isolate the effects of
extinction.  This technique was first used by Whiteoak (1966) to
analyze optical spectrophotometry and has been applied in one form or
another many times since.  We refer to it as
``extinction-without-standards'',
since it does not rely upon a set of unreddened standard
stars to determine the extinction curve.  In this method, a model
atmosphere of the reddened star is determined from its photometric or
spectral properties.  The advantage of this approach is that, in
principle, a perfect, unreddened match can be determined for the
intrinsic SED of the reddened star, eliminating the mismatch error
which plagues pair method curves.  The apparent disadvantages of the
approach are not actual disadvantages, but rather requirements, which
can limit its accuracy and range of applicability.  The first
requirement is that a set of models must exist whose accuracy can be
quantified and validated.  The second is that the observations must
contain adequate ``reddening free'' information to accurately determine
the intrinsic SED of a reddened star.  The third is that the fluxes
must be precisely calibrated, i.e., $r_\lambda$ must well determined.

We have been developing the necessary constituents of this method over
the past five years.  We began by demonstrating that the Kurucz (1991)
\atlas\/ models provide faithful representations of the observed UV and
optical SEDs of near main sequence B stars (FM99).
We then showed that a combination of an observed UV SED and optical
photometry provides adequate reddening independent information to
determine {\em both}\/ the appropriate \atlas\/ model for a reddened
star {\em and} a set of parameters (defined by FM90) which determine
the shape of its interstellar extinction curve.  Subsequently, we
refined the calibration of the \iue\/ data, in order to improve the
quality of the fits and the robustness of the physical information
derived from them (MF00).  Next, we verified the physical parameters
derived from the models through applications of the models to eclipsing
binary data, where the results must agree with other constraints (see,
Fitzpatrick et al.\ 2003 and references therein).  Finally, we have
used \hip\/ data of unreddened B stars to derive a consistent
recalibration of optical and NIR photometry (FM05) and, in the process,
once again demonstrated the internal consistency of the models for near
main sequence B stars.  As a result of these efforts, we now have
internally consistent $r_\lambda$ for \iue\/ and optical and NIR
photometry, and are in a position to apply the results and to quantify
the associated errors.  These are the objectives of the current
paper, and are discussed fully in the next section.

\subsection{Normalizing The Curves}
Once color excesses have been determined, we are faced with the
problem of how to compare excesses derived for lines of sight with
different amounts of extinction.  After all, we are interested in the
``shapes'' of the curves, since these may reveal important clues about
the
size distribution and composition of the dust.  This desire to compare
shapes, brings us to the normalization problem.

From a purely mathematical point of view, a straightforward approach
would be to normalize the curves by their norm,
\begin{equation}
  e(\lambda -V) = \frac{E(\lambda -V)}{\sqrt {\sum_\lambda E(\lambda
  -V)^2}}
  \label{norm}
\end{equation}
With appropriate weighting, this normalization could, on average,
minimize the observational error in the normalization factor and reduce
systematic effects.  However, such a normalization would be sensitive
to the strength of narrow features, such as the 2175\AA\/ bump, and
this could mask the overall agreement between curves over the majority
of the wavelength range.  A second approach is to search for some
immutable feature of the curve and normalize by that.  The idea behind
this procedure is that if some aspect of all curves is always the same,
then all curves can be normalized by the strength of this property and
all of the resulting curves will be directly comparable.  This is the
motivation for the $A_V$ normalization adopted by Cardelli, Clayton, \&
Mathis (1989).  They assume that all extinction curves have a very
similar (although not identical) form for $\lambda \gtrsim V$.
However, there are problems with this normalization as well.  In
particular, as pointed out above, $A_V$ is not a directly measured
quantity, but must be derived from IR photometry and requires
assumptions
about the shape of extinction curves at very long wavelengths.  The
shape
of this extinction is often considered to have a universal form, but
this
has been demonstrated for only a relatively small number of sightlines
(Reike \& Lebofsky 1985; Martin \& Whittet 1990).  In addition,
measurements of
extinction in the IR can be compromised because the stellar SEDs become
increasingly faint at long wavelengths and other sources of light, such
as circumstellar emission or scattering from dust in the near stellar
environment, can contaminate the SED measurements.  Furthermore, IR
color excesses are usually small for stars that are detectable in the
UV, so UV curves normalized by quantities derived from IR photometry
may be affected by large normalization errors.  Consequently, the
absolute level of such curves can be poorly defined.  Finally, even
with the advent of the 2MASS data base, there are still many stars
which do not have IR photometry available.

As a result of the complications mentioned above, and since our intent
is to demonstrate how precisely curves can be measured while making a
minimum number of assumptions, we have opted to use the conventional
$E(B-V)$ normalization, as shown in Eq.~\ref{klambda}. While the
interpretation of \ebv\/ as a measure of the ``amount''
of interstellar dust is not straightforward, its widespread
availability, observational precision, and lack of requisite
assumptions make it the best choice for this study.
Nevertheless, we note that it is a simple matter to transform
from one normalization to another, and emphasize that $E(\lambda-V)$ is
actually the basic measurable quantity.


\section{EXTINCTION-WITHOUT-STANDARDS
\label{secEXTINCTION}}
\subsection{Formulation of the Problem}

Our earlier studies (FM99, FM05) have shown that the observed SEDs,
$f_\lambda$, of lightly- or unreddened main sequence B stars can be
modeled very successfully by using a modified form of Eq. \ref{flux1},
namely,
\begin{equation}
f_{\lambda} = F_{\lambda}\, \theta_R^2\, 10^{-0.4 E(B-V) [k(\lambda-V)
+ R(V)]} \;\;.
\label{flux2}
\end{equation}
The use of absolutely calibrated datasets eliminates the calibration
term $r_\lambda$ and the total extinction $A_\lambda$ has been broken
down
into a normalized shape term ($k({\lambda-V})$), a normalized zeropoint
($R(V) \equiv A_V/\ebv$), and a scale factor (\ebv).  Providing that
the righthand side of the equation can be represented in a
parameterized form, the equation can be treated as a non-linear least
squares problem and the optimal values of the parameters --- which
provide the best fit to the observed SED $f_\lambda$ --- can be derived
along with error estimates.  Because the stars under study were lightly
reddened, we could replace the extinction curve $k({\lambda-V})$ and
the offset term $R(V)$ with average Galactic values without loss of
accuracy.  Using the Kurucz \atlas\/ stellar atmosphere models to
represent $F_\lambda$, the results of the fitting procedure were
estimates of 6 parameters:  \ebv, $\theta_R$, and the four parameters
that define the best-fitting model, i.e., \teff, $\log g$, the
metallicity [m/H], and the microturbulence velocity $v_t$.  We
performed the fits using the MPFIT procedure developed by Craig
Markwardt \footnotemark \footnotetext{Markwardt IDL Library available at
http://astrog.physics.wisc.edu/$\sim$craigm/idl/idl.html.}

The fitting process described above begins to break down when the color
excess \ebv\/ of the target stars exceed $\sim$0.05 mag.  By this we
mean that large systematic residuals begin to appear, which greatly
exceed the measurement errors.  The reason is simple: the wavelength
dependence of interstellar extinction curves varies greatly from
sightline-to-sightline, and once $E(B-V) \gtrsim 0.05$ mag the
differences between the true shapes of the curves and the assumed mean
form begin to exceed the measurement error.  However, FM99 noted that
the SEDs of significantly reddened stars could still be modeled using
Equation~(\ref{flux2}) (and the non-linear least squares approach) if
the wavelength dependence of the extinction curve could be represented
in a flexible form whose shape could be adjusted parametrically to
achieve a best fit to the observations, and if these parameters were
determined {\it simultaneously} with the stellar parameters.  This is
the essence of our ``extinction-without-standards'' approach.

Successfully modeling the shape of reddened stellar SEDs requires four
principal ingredients: 1) an observed SED that spans as large a
wavelength range as possible, 2) an accurate absolute flux calibration
($r_\lambda$ or $C_\lambda$), 3) an extinction curve whose shape can be
described by a manageable set of parameters, and 4) a grid of stellar
surface fluxes, $F_{\lambda}$, whose defining parameters can be
determined from the observational data.  In \S\ref{secDATA} we will
describe the particular datasets used in this paper to demonstrate our
technique.  We have already discussed how MF00 and FM05 have determined
the necessary calibrations.  In the remainder of this section, we
describe our flexible form for the interstellar extinction curve
(\S\ref{secFLEXI}) and the grid of stellar surface fluxes with which we
have developed our approach (\S\ref{secMODELS}).

\subsection{A Flexible Representation of the Interstellar Extinction
Curve
\label{secFLEXI}}

We adopt a flexible and adjustable form for the UV-through-IR extinction
curve, whose shape can be optimized to fit the SED of a reddened star
through the adjustment of a specific set of parameters in the
least-squares minimization procedure.  This curve is illustrated in
Figure \ref{figFLEXI}.  It consists of two main regions: 1) the UV
($\lambda < 2700$ \AA; solid curve) where the parameterized form of
FM90 is adopted and 2) the near-UV/optical/IR ($\lambda > 2700$ \AA;
dashed line) where we use a cubic spline interpolation through a set of
UV (U$_1$, U$_2$), optical (O$_1$, O$_2$, O$_3$, O$_4$), and IR
(I$_1$, I$_2$, I$_3$, I$_4$) anchor points to represent the curve.
The interpolation is performed using the {\em Interactive Data Language
(IDL)} procedure SPLINE.  We adopt a spline representation for the
near-UV/optical/IR curve simply because we do not have reliable,
detailed
information on the wavelength dependence of the extinction in the
near-IR
through near-UV region (1$\mu$m -- 3000 \AA).  It is ironic that the
portion of the curve that is accessible from the ground is more poorly
characterized than the portion accessible only from space.  As a result,
we do not know whether the optical to near-IR region of the curve can be
represented by a compact analytical formula.  Our hope is that, by
applying our procedure to a large sample of sightlines, we will
ultimately be able to characterize the shape of the extinction law in
this region by simple relations and determine whether
sightline-to-sightline variations are correlated with other aspects of
the curve or with interstellar environment. The placement of the spline
points resulted from considerable experimentation, but certainly cannot
be represented as an objectively determined optimal result.  The current
arrangement does, however, allow us to model the major available
datasets
to a level consistent with the observational errors.

The FM90 parameterization scheme contains 6 free parameters to represent
3 functionally separate features that are summed to produce the UV
curve.  An underlying linear component, indicated by the dotted line in
Figure \ref{figFLEXI}, is specified by an intercept $c_1$ and a slope
$c_2$.  The Lorentzian-like 2175 \AA\/ bump is fit by a Drude profile
$D(x,x_0, \gamma)$, where $x_0$ and $\gamma$ specify the position and
FWHM of the bump, respectively, whose strength is determined by a scale
factor $c_3$. Finally, the degree
of departure of the curve in the far-UV from the underlying linear
component is specified by a single parameter $c_4$.  Defining $x \equiv
\lambda^{-1}$, the complete UV function is given by:
\begin{equation}
k(\lambda-V)  = c_1 + c_2 x + c_3 D(x,x_0,\gamma) + c_4 F(x)  \;\; ,
\label{fmfunc}
\end{equation}
where
\begin{equation}
D(x,x_0,\gamma) = \frac{x^2}{(x^2-x_0^2)^2 +x^2\gamma^2} \;\; ,
\label{drude}
\end{equation}
and
{\small
\begin{eqnarray}
F(x; x > 5.9) & = & 0.5392(x-5.9)^2 + 0.05644(x-5.9)^3 \;\; ,  \\
F(x; x \leq 5.9) & = & 0 \;\; .
\label{fuv}
\end{eqnarray}
}
It is worth emphasizing that the FM90 parameterization is a mathematical
scheme only, which allows us to reproduce UV extinction curves in a
shorthand form (and makes the current program possible).  However, it
is not to be assumed that the functional components of the scheme
represent actual separate extinction components arising in distinct
dust grain populations.  This parameterization has proven very useful
and to the best of our knowledge is able to reproduce all known UV
extinction curves to the level of observational error.  Its
flexibility will be illustrated in \S \ref{secRESULTS} below.  Note
that we terminate the FM90 formula at 2700 \AA, although we originally
utilized UV data extending to 3000 \AA.  Additional experience with the
UV data has suggested that real extinction curves begin to exhibit
departures from the FM90 formula near 2700 \AA, due to the unrealistic
extrapolation of the linear component into the blue-violet region.

The 10 spline anchor points which characterize the near-UV/optical/IR
portion of the curve are determined by a least squares fit to the \iue\/
data longward of 2700 \AA\ and the available optical and IR photometry.
Although there are 10 anchor points, there are also 6 constraints, so
the fit actually introduces only 4 additional degrees of freedom.  We
now discuss these constraints in detail.

The two UV anchor points, U$_1$ and U$_2$ at 2700 and 2600 \AA,
respectively, are fixed at the values of the FM90 UV fitting function
at their wavelengths and are not adjustable.  Together with O$_1$ at
3300\AA, these points guarantee that the curve which passes through the
\iue\/ data between 2700 and 3000 \AA\ will join both the UV and
optical portions of the curve smoothly.

The four optical anchor points, O$_1$, O$_2$, O$_3$, and O$_4$ at 3300,
4000, 5530, and 7000 \AA, respectively, are fit under two constraints:
that the interpolated curve produces a value of $k(\lambda-V) = 0$ in
the V band, and that the curve be normalized to unity in \ebv.  Thus,
only two free parameters emerge from this region.

The four IR points, I$_1$, I$_2$, I$_3$, and I$_4$ are located at 0.25,
0.50, 0.75, and 1.0 $\mu$m$^{-1}$, respectively.  These four points are
constrained to satisfy the formula
\begin{equation}
I_n \equiv k(\lambda -V) = k_{IR} \lambda_n^{-1.84} -R(V)
  \label{eqnIR}
\end{equation}
where the scale factor, $k_{IR}$, and the intercept, $R(V)$, are the
only free parameters.  This is the power-law form usually attributed to
IR extinction, with a value for its exponent from Martin \& Whittet
(1990).  The exponent of the power-law could, potentially, be included
as a free parameter in the fitting procedure, and we will investigate
this in the future.  However, our impression is that the IR data
available to us (primarily \tmass\/ $JHK$ photometry) are insufficient
to determine this quantity accurately.  All results presented in this
paper assume an exponent of -1.84 in Equation \ref{eqnIR}.

\subsection{The Stellar Surface Fluxes
\label{secMODELS}}

To represent the intrinsic surface fluxes, $F_\lambda$, of reddened
stars we utilize R.L. Kurucz's line-blanketed, hydrostatic, LTE,
plane-parallel {\em ATLAS9}\/ models, computed in units of
erg~cm$^{-2}$~sec$^{-1}$~\AA$^{-1}$ and the synthetic photometry derived
from the models by Fitzpatrick \& Massa (2005).  These models are
functions of four parameters: \teff\/, \logg\/, [m/H], and $v_t$.
All of these parameters can be
determined in the fitting process although, because of data quality, it
is sometimes necessary to constrain one or more to a reasonable value
and
solve for the others.

The general technique of deriving extinction curves via stellar
atmospheres is, of course, not dependent on the specific set of models
used.  In the present case, the most important consideration for our
adoption of the {\em ATLAS9} models is that they work --- at least
within a specific spectral domain.  FM99 and  FM05 have shown that
these models provide excellent fits to the observed SEDs for lightly-
or un-reddened main sequence B stars throughout the UV through near-IR
spectral region.  In addition, experience with eclipsing binaries (see
Fitzpatrick et al. 2003 and references within) has shown that the good
SED fits also yield accurate estimates of the physical properties of
the stars.  Because of the physical ingredients of the models
(specifically LTE and plane-parallelism) we currently restrict our
attention to the main sequence B stars.  We plan to investigate how
well these models reproduce the SEDs and properties of somewhat more
luminous B stars and also the later O types.  Also, we will take
advantage of more complex models (e.g., the non-LTE TLUSTY models) as
the available grids expand their parameter ranges.

\subsection{Summary}

To summarize, we model the observed SEDs of reddened near main sequence
B stars by treating equation~(\ref{flux2}) as a non-linear least
squares minimization problem.  As a result we can simultaneously obtain
estimates of the physical properties of a reddened star {\it and} the
shape of the interstellar extinction curve distorting the star's SED.
A total of 16 parameters specify the righthand side of the equation,
including $\theta_R$, $E(B-V)$, four to define $F_{\lambda}$, and ten
to define the shape of the extinction curve.  Depending on limitations
of the available data,
and known properties of the stars or interstellar medium,
any subset of the parameters can be constrained
to predetermined values.


\section{THE DATA
\label{secDATA}}

In the following section, we will illustrate the potential of our
extinction-without-standards technique, utilizing a set of 27
lightly-to-heavily reddened stars.  For this demonstration, and indeed
for extinction determinations in general, the ideal SED dataset would
consist of absolutely-calibrated spectrophotometry spanning the
UV-through-IR spectral regions.  Such data would allow a
straightforward comparison between observations and stellar atmosphere
models (since both are presented in simple flux units) and would
provide the most detailed view of the wavelength dependence of
interstellar extinction.  While a small amount of such data is
available (see, e.g., Fitzpatrick at al. 2003), the largest existing
database of absolutely-calibrated spectrophotometry is the
low-resolution archive of the {\it International Ultraviolet Explorer}
satellite (\iue) which covers the UV region only (1150--3000 \AA).  In
the optical and near-IR, the largest SED databases are photometric in
nature, consisting of Johnson, Str\"{o}mgren, and Geneva photometry in
the optical and {\it 2MASS JHK} photometry in the near IR.  Utilizing
these resources, we can examine the UV region for small scale features
but can only study the broad scale wavelength-dependence of extinction
in the optical and near-IR regions.

We use NEWSIPS \iue\/ data (Nichols \& Linsky 1996) obtained from the
MAST archive at STScI.  These data were corrected for residual
systematic errors and placed onto the \hst/FOS flux scale of Bohlin
(1996) using the corrections and algorithms described by MF00.  This
step is absolutely essential for our program since our ``comparison
stars'' are stellar atmosphere models and systematic errors in the
absolute calibration of the data do not cancel out as in the case of
the pair method.  (The NEWSIPS database is also contaminated by
thermally- and temporally-dependent errors, which would not generally
cancel out in the pair method --- see MF00.) Multiple spectra from each
of \iue{\it 's} wavelength ranges (SWP or LWR and LWP) were combined
using
the NEWSIPS error arrays as weights.  Small aperture data were scaled
to the large aperture data and both trailed and point source data were
included.  Short and long wavelength data were joined at 1978~\AA\ to
form a complete spectrum covering the wavelength range $1150 \leq
\lambda \leq 3000$~\AA.  Data longward of 3000~\AA\ were ignored
because they are typically of low quality and subject to residual
systematic effects.  The \iue\/ data were resampled to match the
wavelength binning of the \atlas\/ model atmosphere calculations in the
wavelength regions of interest.

Mean values of the Johnson {\it UBV}, Str\"{o}mgren {\it uvby}$\beta$,
and Geneva {\it UBB$_1$B$_2$VV$_1$G} photometric magnitudes, colors,
and indices for the program stars were acquired from the Mermilliod
et~al. (1997) archive.  \tmass\/ {\it JHK} magnitudes for all stars,
along with their associated errors, were obtained from the \tmass\/
All-Sky Point Source Catalog at the NASA/IPAC Infrared Science
Archive.  Johnson $V-R$, $R-I$, and {\em JHK} data are also available
for some of the stars, and were obtained from the Mermilliod et~al.
archive.


\section{SOME INITIAL RESULTS
\label{secRESULTS}}

In this section, we demonstrate the potential of our
extinction-without-standards technique, utilizing a set of the 27
reddened stars, listed in Table~1, which fall into three groups: 1)
stars in the open cluster IC~4665; 2) stars with moderate-to-heavy
reddening; and 3) lightly reddened stars in a specific region of the
sky.  These representative examples illustrate the advantages of our
approach and highlight several scientific applications which will be
pursed in future studies, using expanded samples of stars.  In
addition,
they provide confirmation of the error analysis incorporated in our
approach.

For this demonstration sample, the SED fitting procedure was applied
as described above, with the following additional details:
\begin{itemize}
\item The SED data modeled in the fitting procedure include \iue\/ UV
spectrophotometry, the Johnson $V$, $B-V$, and $U-B$ indices, the
Str\"{o}mgren $b-y$, $m_1$, $c_1$, and $\beta$ indices, the Geneva $U-B$,
$V-B$, $B_1-B$, $B_2-B$, $V_1-B$, and $G-B$ indices, and the \tmass\/
{\it JHK} magnitudes. Johnson $V-R$, $R-I$, and {\em JHK} data are also
available for a few of the stars.

\item The optical extinction spline point O$_4$ at 7000~\AA\ is only
well-determined when optical $R$ and $I$ band photometry are available.
Therefore in the examples below, we only solve for O$_4$ in such cases.
For the other stars, the optical portion of the extinction curve is
determined only by the spline points O$_1$, O$_2$, and O$_3$.

\item During the $\chi^2$ minimization, a reddened and
distance-attenuated model was created from the current set of input
parameters, and then synthetic photometry was performed on this model
to produce the photometric indices, which were then compared with
observations.  The synthetic photometry was calibrated as described by
FM05, with the calibration extended to redder colors by us.  The UV
model fluxes and recalibrated \iue\/ fluxes, both in units of
erg~cm$^{-2}$~sec$^{-1}$~\AA$^{-1}$, were compared directly.

\item  The initial weighting factors for the various datasets in the
$\chi^2$ minimization were determined from their observational
uncertainties, i.e., $weight \propto \sigma^{-2}$.  We then scaled the
weights of the optical/near-IR photometry so that their total weight
was equal to that of the \iue\/ UV spectrophotometry, thus balancing
the fit between the two datasets.  This is the procedure adopted by
FM05, except that we include the Str\"{o}mgren $\beta$ index along with
the rest of the optical/near-IR photometry, rather than assigning it
its own (high) weight.  FM05 weighted $\beta$ heavily in recognition
of its value as a surface gravity indicator.  However, we have found
that the temperature-sensitivity of $\beta$ (particularly in the later
B stars) combined with observational errors, can lead to very
unsatisfactory fits when $\beta$ is over-emphasized.  Treating $\beta$
in the same manner as the rest of the photometric indices seems to be
the simplest and most reasonable approach.

\item Because interstellar H~{\sc i} Lyman $\alpha$ absorption in
reddened
stars can have a significant impact on the star's apparent continuum
level far from line center at 1215 \AA, we convolve the profile of this
heavily damped line with the model atmosphere SEDs before comparing
with observations.  Along heavily reddened sightlines, where the H~{\sc
i}
column density $N($H~{\sc i}$)$ is high and the signature of the atomic
absorption strong, the value of $N($H~{\sc i}$)$ can actually be
incorporated into the fitting procedure as a free parameter to be
optimized.  We will utilize this capability in future studies.  For the
present, mostly lightly-reddened, sample, however, the $N($H~{\sc i}$)$
values used to construct the line profiles were taken from the
survey of Diplas \& Savage (1994) or else computed from the general
relation $N($H~{\sc i}$) = 4.8\times10^{21} E(B-V)\; {\rm cm}^{-2}$ from
Bohlin, Savage, \& Drake (1978). The inclusion of the Lyman $\alpha$
line insures that we distinguish the effects of dust extinction from
atomic absorption in the far UV region.

\item The uncertainties in the best-fit parameters were determined by
running 50 Monte Carlo simulations for each star, during which the
input data were randomly varied assuming a Gaussian distribution of
observational uncertainties and a new fit performed.  The zero points
and random photometric uncertainties of the short-wavelength and
long-wavelength \iue\/ fluxes were varied as described in FM04;  the
assumed observational errors in the Johnson, Str\"{o}mgren, and Geneva
indices were as given in Table 7 of FM04; and the uncertainties in the
\tmass\/ data were as obtained from the \tmass\/ archive.  In addition,
the $V$ magnitude was assumed to have a 1-$\sigma$ uncertainty of 0.015
mag.  The adopted 1-$\sigma$ uncertainties for each parameter were
taken as the standard deviation of the values produced by the 50
simulation.

\end{itemize}

\subsection{The Open Cluster IC~4665 \label{secIC4665}}

We begin our examples by examining the extinction towards an open
cluster.  While multiple scientific rewards can result from the study
of extinction towards cluster stars (see the discussion in \S
\ref{secDISCUSS}), our primary interest in IC~4665 is to demonstrate
the ``technical'' advantages of our approach.  Namely, the use of
cluster extinction curves to help evaluate the magnitude of the
uncertainties in the measurement of extinction curves, as discussed in
detail by Massa \& Fitzpatrick (1986).  In particular, and because of
its low $E(B-V)$ and late-B stellar population, IC~4665 extinction
curves provide an especially sensitive test of the precision and range
of our extinction-without-standards approach.

The wavelength dependence of extinction towards IC~4655 was first
examined by Hackwell, Hecht, \& Tapia (1991; hereafter HHT) for the
purpose of studying the relationship between extinction and IR
emission, as measured by the {\it Infrared Astronomical Satellite}
(\iras).  This remains one of the most challenging extinction studies
yet performed for two reasons: 1) the mean reddening in the cluster is
very low, $\langle E(B-V) \rangle < 0.2$ mag, and 2) the spectral types
of the available targets run from mid- to late B.  Both facts
exacerbate errors in the standard pair method approach, as has been
shown in Figure~\ref{figMISMATCH}.  HHT recognized these uncertainties
and ultimately concluded that the wavelength dependence of extinction
among the cluster stars is uniform to within their ability to measure
it.

Figure~\ref{figSED1} shows the results of the SED fits for the nine
IC~4665 stars considered here.  The SED's of the best-fitting, reddened
models are shown by the histogram-style curves.  In the UV, the binned
\iue\/ fluxes are shown by the small circles.  In the optical region,
Johnson {\it UBVRI} magnitudes (converted to flux) are indicated by
circles, Str\"{o}mgren {\it uvby} magnitudes by triangles, and Geneva
{\it UB$_1$B$_2$VV$_1$G} magnitudes by diamonds.  In the near-IR,
\tmass\/ and Johnson {\em JHK} magnitudes are shown by the large filled
and open circles, respectively. Note that the photometric data have been
converted to flux units for display purposes only.  The comparison
between models and observations was performed in the native photometric
format (i.e., in magnitudes or colors as noted above).

Figure~\ref{figEXT1} shows our extinction-without-standards curves for
the IC~4665 stars.  The symbols show the actual normalized ratios
between
the models and the stellar SEDs, while the
thick solid curves show the flexible UV-through-IR extinction curves
whose shapes were determined by the fitting procedure.  The curves have
been arbitrarily shifted vertically for clarity, but all are shown
compared with a similarly-shifted estimate of the average Galactic
extinction curve for reference (thick dash-dot curves corresponding to
$R(V)=3.1$, from Fitzpatrick 1999).  As will be discussed further
below, we assumed a value of $R(V) = 3.1$ towards the cluster and did
not
include the IR {\em JHK} data in the fitting procedure.  Thus, only the
average Galactic curve is shown for wavelengths longward of 6000 \AA.

The various parameters determined by the fits are listed in Tables 1
and 2.  The flexible extinction curves themselves, in the form
$E(\lambda-V)/E(B-V)$ can be reconstructed from the parameters
given in Table 2.  The 1-$\sigma$ uncertainties of the extinction
curves are indicated in Figure~\ref{figEXT1} by the grey shaded
regions.  The regions are centered on the means of the 50 Monte Carlo
simulations with which we performed our error analysis and their
thickness shows the standard deviation of the individual simulations.

Figure \ref{figEXT1comp} compares our new
curves in the UV with those derived by HHT using the standard pair
method.  HHT's curves were reconstructed from the data in their Table
5.  Note that the HHT study originally included 17 stars.  We have
eliminated 4 A-type stars, 3 chemically peculiar B-type stars, and 1
B-type shell system from consideration here since their extinction
curves are particularly uncertain.  Thus, Figures~\ref{figEXT1} and
\ref{figEXT1comp} show only the best-determined curves in the cluster,
from the point of view of both the pair method and our technique.  The
remarkable aspect of Figure~\ref{figEXT1comp} is not the scatter among
HHT's curves --- it is exactly what should be expected given the
limitations of the pair method --- but rather the impressive
improvement in precision gained by the extinction-without-standards
technique, as evidenced by the decrease in the curve-to-curve scatter.
Clearly, this higher level of precision allows much firmer conclusions
to be drawn about the degree of intrinsic variation of extinction
across the face of the cluster,
and also affords an improved potential to detect correlated behavior
between the extinction properties and other aspects of the interstellar
environment.

While we will present a full scientific analysis of the IC~4665 results
in a future paper, the issue of the intrinsic variation of extinction
among the cluster stars is important to consider here, since it can shed
light on the accuracy of our error analysis.  The top curve in
Figure~\ref{figEXT1} shows the mean of the nine individual IC~4665
curves.  The error bars show the actual sample standard deviation and
the average Galactic curve is again presented for comparison.  The
critical result is that, over most of its wavelength range, the
standard deviation of the sample about the mean curve is comparable to
the predicted uncertainties in the individual curves, as shown by the
shaded regions.  This indicates that the (small) curve-to-curve
variations seen are at a level consistent with our expected
uncertainties and that the intrinsic level of variation among the
cluster stars must be very low.

Another way to approach the issue of variability is to look at the
scatter among the various parameters which define the extinction
curves.  These are shown in Table 3, where columns 2 and 3 list the
weighted mean values of the parameters and their observed standard
deviations, respectively.  The predicted uncertainties, i.e., the RMS
of the Monto Carlo-based errors for the individual stars, are listed in
column 4.  If our error analysis is reasonable, then the observed
scatter should be the quadratic sum of the expected errors
plus any intrinsic variability.  The value of examining cluster
extinction curves is that one might reasonably suppose (at least as a
starting point) that the individual curves, derived for nearly
coincident lines of sight, are actually
independent measurements of a single ``cluster curve,'' i.e., no
intrinsic scatter.  In such a case, the measured scatter actually
reflects the real measurement errors and provides an important test of
the error analysis.  Examination of Table 3 shows that the predicted
and observed scatters are indeed generally very similar, supporting the
position that any intrinsic variations among the IC~4665 curves are
close to the level of our ability to measure them {\em and} that we
have accurately assessed the uncertainties in our results.  The final
column of Table 3 shows the implied values of the possible small
intrinsic variations.

Several of the individual parameters merit some additional comment.
Both the 2175 \AA\/ bump FWHM ($\gamma$) and its strength ($c_3$ or
$A_{bump}$) show evidence for some small level of variability within the
sample, above our expected measurement errors.  However, this is not
clearcut because these measurements involve the region of the {\em IUE}
spectra which typically has the lowest quality data and it is possible
that weak systematic effects in the data  themselves --- which are not
accounted for in the error analysis --- could produce the small level
of variability seen. One such systematic is a ``reciprocity effect'' in
long wavelength data which is not corrected by the MF00 algorithms (see
Figures 12 and 13 of MF00).  We conclude conservatively that there is
marginal evidence for bump variations within the IC~4665 sample, but
will ultimately rely on studies of several open clusters to determine
whether our error analysis faithfully predict the real measurement
errors in the bump region.

The case of the far-UV curvature (parameter $c_4$), for which the
implied intrinsic scatter is three times greater than our measurement
errors, is more interesting.  It is clear from Figures \ref{figEXT1}
and \ref{figEXT1comp} and Table 3 that most of this apparent
variability arises from two sightlines, namely, those towards HD~161165
and HD~161184.  In fact, the observed scatter in $c_4$ towards the
other seven sightlines is essentially identical to the expected
measurement error.  Because HD~161165 and HD~161184 are the two coolest
stars in the sample, and the only stars with effective temperatures
less than 12,000~K, we suspected that the high $c_4$ values in their
extinction curves might be artifacts of the analysis, resulting from a
failure of the models to accurately portray the far-UV SEDs of these
stars. The investigation presented below suggests that this could well
be the case, with the extinction curve anomalies possibly arising from a
difference between the chemical composition profiles of the stars and
the atmosphere models.

The data in Table 1 show a very uniform ``metallicity'' [m/H] for the
cluster, with a weighted mean of -0.50 and a standard deviation of
0.04.  This is close to, and actually slightly smaller than, our
estimate of the measurement errors, simultaneously confirming the
accuracy of our error analysis and imposing a small upper limit on the
intrinsic compositional variability within the cluster.  Although our
[m/H] values are simple scale factors which apply to a template set of
\atlas\/ solar abundances, FM99 showed that the [m/H] derived in our
analysis is
most sensitive to the abundance of Fe --- due to a very strong opacity
signature in the mid-UV --- and is most analogous to [Fe/H].  Since the
Fe abundance in the \atlas\/ models is $\sim$0.2 dex larger than the
currently accepted solar value of 7.45 (where H = 12.00; Asplund,
Grevesse, \& Sauval 2005), our results suggest that the IC~4665 stars
are deficient by about a factor of two in Fe as compared to the Sun.
The small scatter observed in [m/H] is both satisfying and expected,
since the surface composition of Fe is not subject to evolutionary
modification in these young stars, which presumably all formed from the
same parent material.

We experimented with forcing [m/H] to a more solar-like level for the
IC~4665 stars and immediately found two effects: 1) the $\chi^2$ values
all increased significantly because the Fe features in the mid-UV could
not be fit as well, and 2) most of the extinction curves remained
unchanged, but the anomaly in the far-UV region for HD~161165 and
HD~161184 decreased dramatically. The first result was expected. The
second was a
surprise, but is understandable. If the cluster stars (or at minimum,
the two coolest stars HD~161165 and HD~161184) have a non-solar ratio
of Fe to the light metals, e.g., [C/Fe] $>$ 0, then our best fit models
--- biased towards the Fe abundance --- would underestimate the opacity
due to the light metals.  For the cooler stars, such elements provide
significant opacity in the far-UV region and the \atlas\/ models would
not be able
to account for such a specific opacity difference.
The fitting procedure would respond by finding a higher far-UV
extinction curve.
The curves for the hotter stars are less affected
by changing [m/H] since the light metal opacity is less significant.
We tentatively conclude that the chemical composition of the B stars
in IC~4665 may deviate strongly from that of the Sun, with subsolar Fe
but a more ``normal'' level of the light metals such as C.  This
suggestion is easily tested with a fine analysis of high resolution
stellar spectra and we will pursue this in the future.  On the positive
side, this result suggests that the UV continua of late B and early A
stars might
be
used to determine {\em both} a scaled [m/H] {\em and} a mean [CNO/Fe]
index, assuming a grid of models parameterized by both these composition
indices is available.  On the more sober side, it is a reminder that our
technique is susceptible to stellar abnormalities.  As with the pair
method, all available data should be consulted to determine whether a
particular star is suitable for deriving an extinction curve.

Our final comment  on the IC~4665 results concerns the behavior of the
IR data.  In Figure \ref{figEXT1} we show two sets of {\it JHK}
photometry for each star and for the average cluster curve.  The solid
symbols indicate \tmass\/ measurements and the open symbols show HHT's
data.  The two sets of data systematically differ, with the \tmass\/
results suggesting a value of $R(V)$ somewhat smaller then the average
Galactic value of 3.1 and the HHT data suggesting a slightly larger
value (HHT derive a mean of $R(V) = 3.25$). This latter result is more
consistent with expectations, given that the mean far-UV curve is lower
than the Galactic average (see, e.g., Cardelli et al. 1989), but this
is
insufficient evidence for rejecting the \tmass\/
data.  For now, our solution for this quandary has been to ignore both
sets of data in fitting the IC~4665 SEDs and adopt a default value of
$R(V) = 3.1$.  However, the discrepancy for measurements in this very
complex region bears further investigation, as does that fact that, in
both datasets, the mean curves actually show more extinction in the
2.2~$\mu$m $K$ band than in the 1.65 $\mu$m $H$ band.

The discussion above leads us to three primary conclusions:  1) The
extinction towards IC~4665 shows at most only a small degree of spatial
variability --- comparable with our measurement errors --- and that the
cluster extinction curve would best be represented by averaging the
results for the seven hottest cluster stars studied here; 2) we are
able
to determine accurate [m/H] values from the observations; and 3)
our
analysis yields reliable estimates of the (small) uncertainties in the
extinction-without-standards curves and parameters, barring
the presence of unusual systematic anomalies in the data sample.  The
first and
second
conclusions are scientific issues which we will pursue further in the
future.  The third provides a formidable demonstration of the
superiority of the extinction-without-standards technique over the
classical pair method for deriving extinction curves in both the
precision of the results and the quantification of the uncertainties.

\subsection{Moderately-Reddened Stars \label{secFRIENDS}}

With our error estimates verified, we now examine how curves derived by
the
current approach compare to pair method curves.
Figure~\ref{figSED2} shows the UV-through-IR fits to the SEDs of a set
of
9 moderately-reddened early-B stars, most of which are well known for
their extinction properties, and Figure~\ref{figEXT2} shows the
corresponding extinction curves.  These stars were selected because
they
illustrate the wide range that exists in the wavelength dependence of
both
UV and IR extinction.  For these stars the IR data allow us to determine
the values of $R(V)$ and so the flexible extinction curve fits are
shown
throughout the IR-to-UV domain.  As for the IC~4665 stars, all the
parameters describing the fits are given in Tables 1 and 2.

Also shown in Figure \ref{figEXT2} are UV curves based on the pair
method technique.  The curve for HD~294264 is from Valencic, Clayton,
\& Gordon (2004);  the curves for HD~210121 and HD~27778 are from
unpublished measurements by us; and the others are from the catalog of
FM90.  The agreement between the model-based and pair method curves is
reasonable, and much better than seen for IC~4665.  This is consistent
with spectral mismatch as the prime cause of the existing
discrepancies, since the nine stars have both higher $E(B-V)$ values
and earlier spectral types than the IC~4665 stars, both of which tend
to reduce the influence of mismatch errors.  Note also that the best
agreement between the pair method and the model-based curves occurs for
the star HD~204827, for which identical results are found.  Again, this
is as it should be, since its $E(B-V)$ is the largest of any star in
the sample, minimizing the impact mismatch effects.  The 1-$\sigma$
uncertainties for our flexible extinction curve fits are indicated by
shaded regions around the curves, as in Figure \ref{figEXT1}.

The curves shown in Figure \ref{figEXT2} demonstrate the ability of
the parameterized, flexible extinction curve (see \S\ref{secFLEXI}) to
conform itself to the wide range of extinction curve shapes encountered
in
interstellar space.

 \subsection{Lightly-Reddened Stars \label{secREGIONX}}

Figure~\ref{figSED3} shows a set of nine SED fits for low color excess
stars (0.10 $\leq E(B-V) \leq$ 0.21) located along sightlines bounded
by the Galactic coordinates $347\degr < l < 355\degr$ and $18\degr < b
<
26\degr$.  Figure~\ref{figEXT3} shows the corresponding extinction
curves.
These stars are all mid-to-late B members of the Upper Scorpius complex
(Garrison 1967) and we encountered them while testing a procedure for
scanning the \iue\/ archives and automatically generating model-based
extinction curves for stars in the appropriate spectral range.  When
examining the results for this pilot program which sampled
high-latitude
stars, it became obvious that curves derived from stars in this
specific
region demonstrated similar curve morphology which is distinctly
different
from the Galactic average.  Given the low reddening and preponderance
of
late B stars, this strong systematic behavior would be missed by a pair
method survey, with its large inherent mismatch errors.

Strong regional signatures are important in extinction studies because
they may highlight the effects of specific physical processes on dust
grain populations.  Although we will examine this specific region in the
future, two points are worth mentioning.  First, it is important that
similar curves result from stars with spectral types ranging from B2.5
to B9, verifying that the curves are not the result of some
temperature-dependent systematic effect in fitting process.  It is also
worth noting that the region is near the $\rho$ Oph dark cloud.  The
star $\rho$~Oph~A (HD~147933, whose extinction curve is shown in
Figure~\ref{figEXT2}) is located at a comparable distance and just
south of the region, at Galactic coordinates of ($l$, $b$) =
($353.7\degr, 17.7\degr$).  The mean curve for the nine
lightly-reddened stars is shown at the top of Figure~\ref{figEXT3}, along with the average Galactic curve (dash-dot curve) and the
HD~1479433 curve (dotted curve) for comparison.  In the UV, the
regional curve is seen to be almost identical to that for the more
heavily-reddened HD~147933 ($E(B-V) \simeq 0.5$), suggesting that the
sightlines sample similar dust populations.  Interestingly, however,
the curves are not identical in the IR.  The mean \tmass\/ {\it JHK}
data for the nine star sample implies a value of $R(V) \simeq 3.4$,
slightly larger than the Galactic mean of 3.1, while HD~147933 has a
value of $R(V) \simeq 4.3$.  This contrast between UV and IR behavior
may indicate different physical processes at work along the higher
density sightline towards HD~147933, or perhaps different timescales in
the response of UV and IR extinction to modifications of
dust grain properties.

In producing the extinction curves for these lightly reddened stars, we
found that, in some cases, the value of $k_{IR}$ was very poorly
determined and produced (presumably) spurious ``bumps and wiggles'' in
the IR portion of the curves.  To eliminate this effect, we imposed a
constraint on $k_{IR}$ for the whole sample, namely $k_{IR} =
0.63R(V)-0.84$.  This is taken from F04, who found a very strong
relationship between $R(V)$ and $k_{IR}$ from a larger sample of more
heavily reddened stars (see Figure 6 of F04).  The ability to impose
scientifically reasonable constraints on the fitting procedure is a
major advantage of the extinction-without-standards approach, and can
potentially allow high quality extinction results to be derived from
stars with even lower reddenings than those shown in
Figure~\ref{figEXT3}.


\section{DISCUSSION \label{secDISCUSS}}

In the previous sections we first provided an overview of the process
used to determine an extinction curve, clarifying the measurement
process.  We then presented a new method for deriving UV-through-IR
extinction curves, based on the use of stellar atmosphere models to
provide estimates of the intrinsic (unreddened) stellar SEDs rather
than unreddened (or lightly reddened) ``standard'' stars.  We have shown
that this ``extinction-without-standards'' technique greatly increases
the accuracy of the derived extinction curves, particularly in the cases of
low reddening and cool spectral types (i.e., late-B), and allows a
realistic estimation of the uncertainties.  A side benefit of the
technique is the simultaneous determination of fundamental properties
of
the reddened stars themselves (\teff, \logg, [m/H], and \vturb), making
the procedure valuable for both stellar and interstellar studies. Given
the physical limitations of the \atlas\/ models we currently employ,
the
technique is limited to near main sequence B stars.  However, in
principal, the procedure can be adapted to any class of star for which
accurate model SEDs are available and for which the signature of
interstellar reddening can be distinguished from those of the stellar
parameters.  Although we developed the procedure based on \iue\/
spectrophotometry in the UV, and photometry in the optical and near-IR
(requiring a calibration of synthetic optical and near-IR photometry),
the ideal application of the technique would be with spectrophotometric
data throughout the UV-through-IR domain, allowing the most detailed
examination of the wavelength dependence of the extinction curves.

The specific scientific advantages afforded by the
extinction-without-standards technique can be summarized as follows:

\noindent {\underline{\bf Increased Precision:}} The increased
precision
in the derived extinction curves allows us to improve our understanding
of extinction in a number of ways. First and most simply, we will
determine the basic UV-to-IR wavelength dependence of extinction along
a
wide variety of sightlines more precisely than has been possible in the
past.  Second --- and given that we know that extinction curve
morphology
varies widely from sightline-to-sightline (see Figure \ref{figEXT2}!)
--- we will be able to study the form of the variability and search for
relationships between various features and wavelength domains using a
data set with small and well-defined uncertainties.  Such
relationships,
e.g., the correlation between $R(V)$ and the steepness of UV
extinction discovered by Cardelli et al. (1989), provide important
constraints on the dust grain population causing the extinction.
Non-correlations can be equally important. For example, the lack of a
correlation between the position and width of the 2175 \AA\/ bump
demonstrated by Fitzpatrick \& Massa (1986) remains a strong constraint
on models for the bump carrier (Draine 2003).  In either case, a
precise
knowledge of the measurement errors is required for transforming an
observation into a scientific constraint. Finally, we will be able to
place much stronger and more well-defined limits on the relationship
between extinction curve morphology and interstellar environment.
Curves
derived from lines of sight to specific, localized regions, such as
towards open clusters, can be particularly useful for relating curve
properties to physical processes occurring in the interstellar medium
(see, e.g., the study of Cepheus~OB3 by Massa \& Savage 1984 and
Trumpler~37 by Clayton \& Fitzpatrick 1987).

\noindent {\underline{\bf Access to lightly-reddened sightlines:}} The
ability to accurately probe low $E(B-V)$ sightlines (as exemplified in
Figures \ref{figEXT1} and \ref{figEXT3}) opens the door to studies
of dust in regions that have not been thoroughly explored.  These
include halo dust, dust in very low density regions, and local dust.
Halo dust is especially important since we must contend with its
effects every time we look out of the Galaxy.  There are indications
(Kiszkurno-Koziej \& Lequeux 1987) that its properties differ
systematically from dust at lower Galactic altitudes, and this result
needs to be verified on a star-by-star basis.  The nature of low
density dust provides insights into the processing which occurs in
hostile environments.  Clayton et al.\ (2000) presented observations
of dust from low density sightlines, and their results were intriguing.
However, they were forced to select sightlines which accumulate fairly
substantial color excesses, introducing the possibility of mixed grain
populations.  Furthermore, the results for their least-reddened
sightlines were, as they acknowledge, poorly determined, forcing them
to base their conclusions on a global average of properties of their
sample.  Finally, measuring the properties of local dust is important
because it allows us to search for isolated, relatively homogeneous
environments with uniform extinction curve shapes.  These may signal
physically and kinematically isolated regions which would be ideal for
follow up interstellar line studies.  In addition, {\it Hipparcos}\/
parallax data exist for nearby stars and provide an opportunity to
study
the 3-dimensional structure of local extinction.

\noindent {\underline{\bf Access to the mid-to-late B stars:}} These
stars are especially important because their space density is higher
than that of the early-B stars usually used in extinction studies and
thus their inclusion increases the number of stars available to create
curves for nearby sightlines. Study of the mid-to-late~B stars will
enable the examination of the spatial structure of local dust
absorption more thoroughly than previously possible and will greatly
enhance our understanding of the local interstellar medium.  The
ability to construct accurate curves for mid-to-late B stars will also
allow us to study extinction in open clusters and associations which are
dominated by these stars, such as the Pleiades, $\alpha$~Per cluster,
and IC~4665 (see Figure \ref{figEXT1}).

\noindent {\underline{\bf Automation:}} Because our model atmosphere
approach does not require human intervention, once the basic data have
been assembled, it is possible to process large data sets at one time.
Naturally, the automated results must be inspected for outliers and
data anomalies.  Nevertheless, this approach reduces the work load
considerably and a first-attempt yielded the regional anomaly shown in
Figure \ref{figEXT3}.  Furthermore, since the results are produced
in a uniform manner, it is a relatively simple matter to inspect them
for correlations between various curve properties, for anomalous curve
shapes, and for spatial trends on the sky.

\noindent {\underline{\bf Stellar properties:}} In addition to dust
parameters, our technique provides a meaningful, quantitative physical
properties for the reddened stars.  The temperature and surface gravity
information will be useful for population studies of B stars in the
field
and in clusters.  However, perhaps the most useful property will be the
metallicity.  We have demonstrated that several stars in the same
cluster,
which have a range in temperatures and gravities, all yield the same
[m/H].  This verifies the sensitivity of our fitting procedure to this
important quantity.  As a result, we are confident that application of
our
procedure to large scale surveys of reddened, near main sequence B stars
can provide a census of the distribution of metallicity throughout the
Galaxy and the local universe.


\begin{acknowledgments}
E.F. acknowledges support from NASA grant NAG5-12137, NAG5-10385, and
NNG04GD46G.  D.M. acknowledges support from NASA grant NNG04EC01P.
Some of the data presented in this paper were obtained from the
Multimission Archive at the Space Telescope Science Institute (MAST).
STScI is operated by the Association of Universities for Research in
Astronomy, Inc., under NASA contract NAS5-26555. Support for MAST for
non-HST data is provided by the NASA Office of Space Science via grant
NAG5-7584 and by other grants and contracts.  This publication also
makes use of data products from the Two Micron All Sky Survey, which is
a joint project of the University of Massachusetts and the Infrared
Processing and Analysis Center/California Institute of Technology,
funded by the National Aeronautics and Space Administration and the
National Science Foundation.
\end{acknowledgments}


\bibliographystyle{apj}


\begin{deluxetable}{lllcrcrc}
\tabletypesize{\scriptsize}
\tablewidth{0pc}
\rotate
\tablecaption{Best-Fit Parameters for Stars in Figures 3--9}
\tablehead{
\colhead{Star}              &
\colhead{Spectral}          &
\colhead{$T_{eff}$}         &
\colhead{$\log g$\tablenotemark{a}}          &
\colhead{[m/H]}             &
\colhead{$v_{turb}$\tablenotemark{b}}        &
\colhead{$\theta_R$}        &
\colhead{$E(B-V)$}              \\
\colhead{}                  &
\colhead{Type}              &
\colhead{(K)}               &
\colhead{}                  &
\colhead{}                  &
\colhead{$\rm (km/s)$}      &
\colhead{(mas)}             &
\colhead{(mag)}}
\startdata
\multicolumn{8}{l}{\bf FIGURES 3 \& 4:} \\
HD 161165 &    B8.5 V & $ 11774\pm 106$ & $ 3.79\pm 0.10$ & $-0.53\pm
0.06$ &
$  0.4\pm  0.4$ & $0.0320\pm0.0004$ & $ 0.19\pm 0.00$ \\
HD 161184 &      B8 V & $ 11438\pm 108$ & $ 4.05\pm 0.09$ & $-0.54\pm
0.06$ &
$  1.5\pm  0.3$ & $0.0426\pm0.0005$ & $ 0.16\pm 0.00$ \\
HD 161572 &      B6 V & $ 14968\pm 494$ & $ 3.94\pm 0.13$ & $-0.49\pm
0.07$ &
$0$ & $0.0414\pm0.0009$ & $ 0.15\pm 0.01$ \\
HD 161573 &     B3 IV & $ 16574\pm 511$ & $ 3.93\pm 0.13$ & $-0.40\pm
0.05$ &
$0$ & $0.0563\pm0.0010$ & $ 0.20\pm 0.01$ \\
HD 161603 &     B5 IV & $ 14734\pm 448$ & $ 3.79\pm 0.12$ & $-0.52\pm
0.06$ &
$0$ & $0.0482\pm0.0010$ & $ 0.18\pm 0.01$ \\
HD 161660 &      B7 V & $ 15468\pm 510$ & $ 4.05\pm 0.12$ & $-0.54\pm
0.09$ &
$  2.2\pm  1.0$ & $0.0374\pm0.0007$ & $ 0.16\pm 0.01$ \\
HD 161677 &     B5 IV & $ 15194\pm 406$ & $ 3.63\pm 0.13$ & $-0.50\pm
0.05$ &
$0$ & $0.0527\pm0.0009$ & $ 0.19\pm 0.01$ \\
HD 161734 &      B8 V & $ 13125\pm 292$ & $ 3.47\pm 0.12$ & $-0.53\pm
0.06$ &
$0$ & $0.0294\pm0.0004$ & $ 0.25\pm 0.01$ \\
HD 162028 &      B6 V & $ 14056\pm 299$ & $ 4.05\pm 0.10$ & $-0.49\pm
0.06$ &
$0$ & $0.0451\pm0.0007$ & $ 0.15\pm 0.01$ \\
\multicolumn{8}{l}{\bf FIGURES 6 \& 7:} \\
 HD 21483 &    B3 III & $ 19741\pm 979$ & $ 3.55\pm 0.54$ & $-0.25\pm
 0.15$ &
$0$ & $0.0710\pm0.0018$ & $ 0.56\pm 0.01$ \\
 HD 27778 &      B4 V & $ 17176\pm 492$ & $ 4.05\pm 0.14$ & $-0.33\pm
 0.04$ &
$0$ & $0.0815\pm0.0018$ & $ 0.37\pm 0.01$ \\
 HD 37061 &      B1 V & $ 30734\pm 452$ & $4.3$ & $-0.64\pm 0.04$ & $8$
 &
$0.0804\pm0.0012$ & $ 0.56\pm 0.00$ \\
 HD 37367 &   B2 IV-V & $ 19850\pm 603$ & $ 3.56\pm 0.19$ & $-0.29\pm
 0.09$ &
$  3.5\pm  1.2$ & $0.0961\pm0.0021$ & $ 0.41\pm 0.01$ \\
 HD 37903 &    B1.5 V & $ 23677\pm 907$ & $4.3$ & $-0.28\pm 0.05$ & $0$
 &
$0.0389\pm0.0010$ & $ 0.37\pm 0.01$ \\
HD 147933 &     B2 IV & $ 25032\pm1098$ & $4.3$ & $-0.32\pm 0.07$ & $0$
&
$0.2280\pm0.0067$ & $ 0.49\pm 0.01$ \\
HD 204827 &      B0 V & $ 31063\pm 390$ & $4.3$ & $-0.65\pm 0.08$ & $8$
&
$0.0508\pm0.0011$ & $ 1.09\pm 0.01$ \\
HD 210121 &      B3 V & $ 18024\pm1079$ & $ 3.26\pm 0.47$ & $-0.74\pm
0.09$ &
$0$ & $0.0405\pm0.0012$ & $ 0.38\pm 0.01$ \\
HD 294264 &     B3 Vn & $ 19830\pm 802$ & $4.3$ & $-0.54\pm 0.08$ & $0$
&
$0.0412\pm0.0010$ & $ 0.52\pm 0.01$ \\
\multicolumn{8}{l}{\bf FIGURES 8 \& 9:} \\
HD 142096 &    B2.5 V & $ 18039\pm 482$ & $4.3$ & $-0.11\pm 0.05$ & $0$
&
$0.1250\pm0.0021$ & $ 0.17\pm 0.01$ \\
HD 142165 &    B6 IVn & $ 14402\pm 287$ & $4.3$ & $-0.70\pm 0.06$ & $0$
&
$0.1144\pm0.0014$ & $ 0.13\pm 0.01$ \\
HD 142315 &    B8 Vnn & $ 12009\pm 108$ & $ 4.24\pm 0.08$ & $-0.76\pm
0.07$ &
$  1.8\pm  0.4$ & $0.0692\pm0.0009$ & $ 0.13\pm 0.00$ \\
HD 142378 &      B3 V & $ 17434\pm 712$ & $4.3$ & $-0.21\pm 0.17$ &
$  0.9\pm  2.6$ & $0.0828\pm0.0018$ & $ 0.17\pm 0.01$ \\
HD 143567 &     B9 Va & $ 11226\pm  43$ & $4.3$ & $-0.54\pm 0.04$ & $0$
&
$0.0636\pm0.0007$ & $ 0.16\pm 0.00$ \\
HD 145554 &    B9 Van & $ 11203\pm  75$ & $ 4.29\pm 0.06$ & $-0.71\pm
0.06$ &
$0$ & $0.0600\pm0.0009$ & $ 0.21\pm 0.00$ \\
HD 146001 &     B7 IV & $ 13922\pm 256$ & $4.3$ & $-0.68\pm 0.08$ &
$  1.4\pm  0.6$ & $0.0945\pm0.0015$ & $ 0.17\pm 0.01$ \\
HD 146029 &     B9 Va & $ 10790\pm  68$ & $ 4.11\pm 0.08$ & $-0.84\pm
0.07$ &
$  1.5\pm  0.4$ & $0.0619\pm0.0007$ & $ 0.14\pm 0.00$ \\
HD 146416 &    B9 Vnn & $ 11394\pm  82$ & $ 4.13\pm 0.08$ & $-0.53\pm
0.04$ &
$0$ & $0.0756\pm0.0009$ & $ 0.10\pm 0.00$ \\
\enddata
\tablenotetext{a}{The maximum allowed value of the surface gravity was
constrained to be $(\log g)_{max}$ = 4.3, the approximate ZAMS gravity
for stars in the relevant mass range.  Stars whose best-fit SED models
required this maximum value are indicated by \logg\/ entries of ``4.3",
without error bars.  The uncertainties listed for stars whose best-fit
\logg\/ values are close to this maximum may be somewhat underestimated
since the upper range in the error simulations was truncated in the
same way.}
\tablenotetext{b}{The allowed values of \vturb\/ were constrained to
lie between 0 and 8 \kms, i.e., the range of values available in the
\atlas\/ model grid.  Stars whose best-fit SED models required these
limiting values are indicated by \vturb\/ entries of ``0" or ``8",
without error bars.  The uncertainties for stars with best-fit \vturb\/
values close to these limits may be underestimated due to this
truncation.}
\end{deluxetable}


\begin{deluxetable}{lccccrrcccrccc}
\tabletypesize{\scriptsize}
\tablewidth{0pc}
\rotate
\tablecaption{Best-Fit Extinction Curve Parameters for Stars in Figures
3--9}
\tablehead{
\colhead{}                  &
\multicolumn{2}{c}{IR Coefficients\tablenotemark{a}}   &
\colhead{}                  &
\multicolumn{3}{c}{Optical Spline Points\tablenotemark{b}}   &
\colhead{}                  &
\multicolumn{6}{c}{UV Coefficients}        \\ \cline{2-3} \cline{5-7}
\cline{9-14}
\colhead{Star}              &
\colhead{$R(V)$}            &
\colhead{$k_{IR}$}          &
\colhead{}                  &
\colhead{$O_1$}             &
\colhead{$O_2$}             &
\colhead{$O_3$}             &
\colhead{}                  &
\colhead{$x_0$}             &
\colhead{$\gamma$}          &
\colhead{$c_1$}             &
\colhead{$c_2$}             &
\colhead{$c_3$}             &
\colhead{$c_4$}             }
\startdata
\multicolumn{13}{l}{\bf FIGURES 3 \& 4:} \\
HD 161165 & \nodata & \nodata &  & $ 1.90\pm 0.11$ & $ 1.32\pm 0.01$ &
$-0.00$
 &  & $ 4.54\pm 0.01$ & $ 0.90\pm 0.05$ & $ 0.58\pm 0.41$ & $ 0.50\pm
 0.08$ &
$ 2.80\pm 0.30$ & $ 0.80\pm 0.05$ \\
HD 161184 & \nodata & \nodata &  & $ 1.87\pm 0.18$ & $ 1.32\pm 0.02$ &
$-0.00$
 &  & $ 4.57\pm 0.01$ & $ 0.84\pm 0.06$ & $ 0.63\pm 0.44$ & $ 0.47\pm
 0.07$ &
$ 2.43\pm 0.28$ & $ 0.93\pm 0.07$ \\
HD 161572 & \nodata & \nodata &  & $ 1.53\pm 0.35$ & $ 1.27\pm 0.03$ &
$-0.01$
 &  & $ 4.59\pm 0.01$ & $ 0.84\pm 0.04$ & $ 0.69\pm 0.52$ & $ 0.40\pm
 0.10$ &
$ 3.09\pm 0.29$ & $ 0.50\pm 0.07$ \\
HD 161573 & \nodata & \nodata &  & $ 1.65\pm 0.21$ & $ 1.29\pm 0.02$ &
$-0.01$
 &  & $ 4.58\pm 0.01$ & $ 0.86\pm 0.06$ & $ 0.65\pm 0.40$ & $ 0.36\pm
 0.09$ &
$ 2.93\pm 0.33$ & $ 0.41\pm 0.04$ \\
HD 161603 & \nodata & \nodata &  & $ 1.67\pm 0.27$ & $ 1.29\pm 0.02$ &
$-0.01$
 &  & $ 4.57\pm 0.01$ & $ 0.88\pm 0.04$ & $ 0.77\pm 0.38$ & $ 0.33\pm
 0.08$ &
$ 3.31\pm 0.27$ & $ 0.42\pm 0.04$ \\
HD 161660 & \nodata & \nodata &  & $ 1.30\pm 0.31$ & $ 1.25\pm 0.03$ &
$-0.01$
 &  & $ 4.57\pm 0.01$ & $ 1.09\pm 0.07$ & $-0.74\pm 0.50$ & $ 0.69\pm
 0.09$ &
$ 4.78\pm 0.70$ & $ 0.41\pm 0.05$ \\
HD 161677 & \nodata & \nodata &  & $ 1.78\pm 0.23$ & $ 1.30\pm 0.02$ &
$-0.01$
 &  & $ 4.56\pm 0.01$ & $ 0.95\pm 0.04$ & $ 0.29\pm 0.42$ & $ 0.48\pm
 0.08$ &
$ 3.47\pm 0.31$ & $ 0.41\pm 0.05$ \\
HD 161734 & \nodata & \nodata &  & $ 2.03\pm 0.18$ & $ 1.33\pm 0.02$ &
$-0.00$
 &  & $ 4.57\pm 0.01$ & $ 1.06\pm 0.03$ & $ 0.12\pm 0.25$ & $ 0.54\pm
 0.05$ &
$ 4.28\pm 0.24$ & $ 0.48\pm 0.03$ \\
HD 162028 & \nodata & \nodata &  & $ 1.81\pm 0.27$ & $ 1.30\pm 0.02$ &
$-0.01$
 &  & $ 4.57\pm 0.01$ & $ 0.85\pm 0.05$ & $ 0.60\pm 0.47$ & $ 0.43\pm
 0.10$ &
$ 3.19\pm 0.29$ & $ 0.55\pm 0.06$ \\
\multicolumn{13}{l}{\bf FIGURES 6 \& 7:} \\
 HD 21483 & $ 2.90\pm 0.06$ & $ 0.98\pm 0.09$ &  & $ 2.25\pm 0.08$ &
$ 1.35\pm 0.01$ & $ 0.00$ &  & $ 4.62\pm 0.01$ & $ 1.10\pm 0.04$ &
$-0.41\pm 0.24$ & $ 0.89\pm 0.06$ & $ 2.85\pm 0.23$ & $ 0.57\pm 0.03$ \\
 HD 27778 & $ 2.63\pm 0.09$ & $ 0.79\pm 0.13$ &  & $ 2.22\pm 0.10$ &
$ 1.34\pm 0.01$ & $-0.00$ &  & $ 4.61\pm 0.01$ & $ 1.18\pm 0.03$ &
$-0.79\pm 0.19$ & $ 0.94\pm 0.04$ & $ 3.57\pm 0.19$ & $ 0.72\pm 0.02$ \\
 HD 37061 & $ 4.51\pm 0.06$ & $ 2.11\pm 0.07$ &  & $ 1.93\pm 0.03$ &
$ 1.33\pm 0.01$ & $-0.00$ &  & $ 4.58\pm 0.00$ & $ 0.89\pm 0.02$ &
$ 1.74\pm 0.16$ & $ 0.16\pm 0.02$ & $ 1.34\pm 0.06$ & $ 0.05\pm 0.01$ \\
 HD 37367 & $ 2.98\pm 0.10$ & $ 0.95\pm 0.15$ &  & $ 2.08\pm 0.06$ &
$ 1.31\pm 0.01$ & $ 0.00$ &  & $ 4.60\pm 0.00$ & $ 0.83\pm 0.01$ &
$ 1.37\pm 0.27$ & $ 0.41\pm 0.05$ & $ 2.98\pm 0.08$ & $ 0.28\pm 0.03$ \\
 HD 37903 & $ 3.85\pm 0.10$ & $ 1.68\pm 0.13$ &  & $ 2.01\pm 0.09$ &
$ 1.33\pm 0.01$ & $-0.01$ &  & $ 4.63\pm 0.00$ & $ 0.97\pm 0.04$ &
$ 1.27\pm 0.24$ & $ 0.35\pm 0.05$ & $ 2.25\pm 0.16$ & $ 0.49\pm 0.03$ \\
HD 147933 & $ 4.30\pm 0.08$ & $ 1.92\pm 0.10$ &  & $ 1.97\pm 0.07$ &
$ 1.32\pm 0.01$ & $ 0.00$ &  & $ 4.57\pm 0.00$ & $ 0.97\pm 0.02$ &
$ 1.43\pm 0.20$ & $ 0.23\pm 0.04$ & $ 3.09\pm 0.12$ & $ 0.33\pm 0.02$ \\
HD 204827 & $ 2.45\pm 0.05$ & $ 0.75\pm 0.07$ &  & $ 2.17\pm 0.03$ &
$ 1.37\pm 0.01$ & $ 0.01$ &  & $ 4.62\pm 0.00$ & $ 0.99\pm 0.02$ &
$-1.55\pm 0.10$ & $ 1.25\pm 0.02$ & $ 2.68\pm 0.07$ & $ 0.85\pm 0.03$ \\
HD 210121 & $ 2.19\pm 0.11$ & $ 0.67\pm 0.14$ &  & $ 2.76\pm 0.11$ &
$ 1.40\pm 0.01$ & $-0.01$ &  & $ 4.52\pm 0.01$ & $ 1.35\pm 0.08$ &
$-3.05\pm 0.29$ & $ 1.78\pm 0.08$ & $ 3.41\pm 0.48$ & $ 0.76\pm 0.02$ \\
HD 294264 & $ 5.55\pm 0.08$ & $ 2.87\pm 0.10$ &  & $ 1.85\pm 0.08$ &
$ 1.31\pm 0.01$ & $ 0.01$ &  & $ 4.57\pm 0.01$ & $ 1.02\pm 0.04$ &
$ 1.62\pm 0.14$ & $ 0.08\pm 0.04$ & $ 2.07\pm 0.15$ & $ 0.24\pm 0.01$ \\
\multicolumn{13}{l}{\bf FIGURES 8 \& 9:} \\
HD 142096 & $ 3.82\pm 0.24$ & $ 1.56\pm 0.15$ &  & $ 2.02\pm 0.17$ &
$ 1.31\pm 0.02$ & $-0.00$ &  & $ 4.59\pm 0.01$ & $ 0.87\pm 0.03$ &
$ 1.23\pm 0.27$ & $ 0.29\pm 0.07$ & $ 2.47\pm 0.16$ & $ 0.07\pm 0.03$ \\
HD 142165 & $ 3.08\pm 0.26$ & $ 1.10\pm 0.16$ &  & $ 1.87\pm 0.30$ &
$ 1.30\pm 0.03$ & $-0.01$ &  & $ 4.61\pm 0.01$ & $ 0.85\pm 0.04$ &
$ 0.73\pm 0.40$ & $ 0.49\pm 0.07$ & $ 2.63\pm 0.22$ & $ 0.23\pm 0.05$ \\
HD 142315 & $ 3.37\pm 0.34$ & $ 1.28\pm 0.21$ &  & $ 1.96\pm 0.18$ &
$ 1.32\pm 0.02$ & $-0.00$ &  & $ 4.57\pm 0.02$ & $ 0.85\pm 0.07$ &
$ 0.98\pm 0.50$ & $ 0.38\pm 0.09$ & $ 2.66\pm 0.40$ & $ 0.29\pm 0.07$ \\
HD 142378 & $ 3.64\pm 0.18$ & $ 1.45\pm 0.11$ &  & $ 2.28\pm 0.25$ &
$ 1.33\pm 0.02$ & $-0.00$ &  & $ 4.64\pm 0.02$ & $ 0.63\pm 0.08$ &
$ 1.51\pm 0.40$ & $ 0.41\pm 0.08$ & $ 0.91\pm 0.22$ & $ 0.09\pm 0.06$ \\
HD 143567 & $ 2.76\pm 0.18$ & $ 0.90\pm 0.12$ &  & $ 1.88\pm 0.08$ &
$ 1.33\pm 0.01$ & $-0.00$ &  & $ 4.59\pm 0.01$ & $ 0.92\pm 0.06$ &
$ 0.12\pm 0.52$ & $ 0.51\pm 0.09$ & $ 3.53\pm 0.45$ & $ 0.24\pm 0.08$ \\
HD 145554 & $ 3.65\pm 0.19$ & $ 1.46\pm 0.12$ &  & $ 1.79\pm 0.10$ &
$ 1.31\pm 0.01$ & $ 0.00$ &  & $ 4.58\pm 0.01$ & $ 0.95\pm 0.05$ &
$ 0.76\pm 0.40$ & $ 0.33\pm 0.07$ & $ 3.54\pm 0.34$ & $ 0.24\pm 0.06$ \\
HD 146001 & $ 3.53\pm 0.22$ & $ 1.38\pm 0.14$ &  & $ 1.69\pm 0.22$ &
$ 1.29\pm 0.02$ & $-0.00$ &  & $ 4.54\pm 0.01$ & $ 1.26\pm 0.15$ &
$ 0.12\pm 0.74$ & $ 0.43\pm 0.11$ & $ 6.00\pm 1.61$ & $ 0.30\pm 0.05$ \\
HD 146029 & $ 3.56\pm 0.27$ & $ 1.40\pm 0.17$ &  & $ 2.11\pm 0.17$ &
$ 1.34\pm 0.01$ & $-0.00$ &  & $ 4.57\pm 0.02$ & $ 0.88\pm 0.07$ &
$ 1.85\pm 0.57$ & $ 0.18\pm 0.11$ & $ 3.30\pm 0.46$ & $ 0.62\pm 0.11$ \\
HD 146416 & $ 2.81\pm 0.36$ & $ 0.93\pm 0.23$ &  & $ 2.05\pm 0.22$ &
$ 1.34\pm 0.02$ & $-0.01$ &  & $ 4.53\pm 0.02$ & $ 0.89\pm 0.08$ &
$ 0.45\pm 0.82$ & $ 0.44\pm 0.16$ & $ 3.08\pm 0.57$ & $ 0.76\pm 0.15$ \\
\enddata
\tablenotetext{a}{Values of $R(V)$ and $k_{IR}$ were not determined for
IC~4665 stars, i.e., the first group of entries below (HD~161165
through HD~162028).  See the discussion in \S 5.1.  For the
lightly-reddened stars in the third group of entries (HD~142096 through
HD~146416), the values of $R(V)$ and $k_{IR}$ were constrained to
follow the relation $k_{IR} = 0.63R(V) - 0.84$ from F04.  See the
discussion in \S 5.3.}
\tablenotetext{b}{The stars HD~37061, HD~37903, HD~147933, and
HD~294264 have Johnson $R$ and $I$ photometry available and thus
allowed us to solve for the fourth optical spline point $O_4$ at 7000
\AA.  The values are: $-0.99\pm0.02$, $-0.92\pm0.04$, $-1.06\pm0.04$,
$-1.16\pm0.03$, respectively.  For the rest of the sample, the optical
spline is determined only by the points $O_1$, $O_2$, and $O_3$ at 3300
\AA, 4000 \AA, and 5530 \AA, respectively.}
\end{deluxetable}


\begin{deluxetable}{lrccc}
\tablewidth{0pc}
\tablecaption{Mean Extinction Parameters for IC~4665}
\tablehead{
\colhead{Parameter}     &
\colhead{Weighted}      &
\colhead{Sample}     &
\colhead{RMS of}        &
\colhead{Implied}        \\
\colhead{}                   &
\colhead{Mean\tablenotemark{a}}               &
\colhead{Standard Deviation\tablenotemark{b}}          &
\colhead{Monte Carlo Errors\tablenotemark{c}}      &
\colhead{Intrinsic Scatter\tablenotemark{d}}}
\startdata
     $O_1$ & $ 1.819$ & $ 0.219$ & $ 0.244$ & \nodata \\
     $O_2$ & $ 1.306$ & $ 0.024$ & $ 0.021$ & $ 0.012$ \\
     $O_3$ & $-0.004$ & $ 0.003$ & $ 0.001$ & $ 0.003$ \\
     $x_0$ & $ 4.569$ & $ 0.015$ & $ 0.011$ & $ 0.010$ \\
  $\gamma$ & $ 0.922$ & $ 0.096$ & $ 0.050$ & $ 0.081$ \\
     $c_1$ & $ 0.380$ & $ 0.475$ & $ 0.428$ & $ 0.206$ \\
     $c_2$ & $ 0.478$ & $ 0.108$ & $ 0.085$ & $ 0.067$ \\
     $c_3$ & $ 3.273$ & $ 0.736$ & $ 0.359$ & $ 0.643$ \\
     $c_4$ & $ 0.500$ & $ 0.191$ & $ 0.052$ & $ 0.184$ \\
$A_{bump}$\tablenotemark{e} & $ 5.633$ & $ 0.709$ & $ 0.319$ & $ 0.634$
\\
\enddata
\tablenotetext{a}{Mean values computed using 1/$\sigma^2$ weighting,
with $\sigma$ values as given in Table 2.}
\tablenotetext{b}{Standard deviation of the nine measurements shown in
Table 2 for each parameter.}
\tablenotetext{c}{RMS value of the nine Monte Carlo-based uncertainties
listed in Table 2 for each parameter.}
\tablenotetext{d}{Computed based on the assumption that the observed
scatter (in the 3rd column) is the quadratic sum of the measurement
errors (in the 4th column) and the intrinsic scatter.}
\tablenotetext{e}{$A_{bump}$ ($\equiv \frac{\pi c_3}{2 \gamma}$) is the
area of the Lorentzian-like 2175 \AA\/ bump for an extinction curve
normalized by ${\rm E(B-V)}$.}

\end{deluxetable}

{\onecolumn

\begin{figure}[ht]
\figurenum{1}
\epsscale{0.85}
\plotone{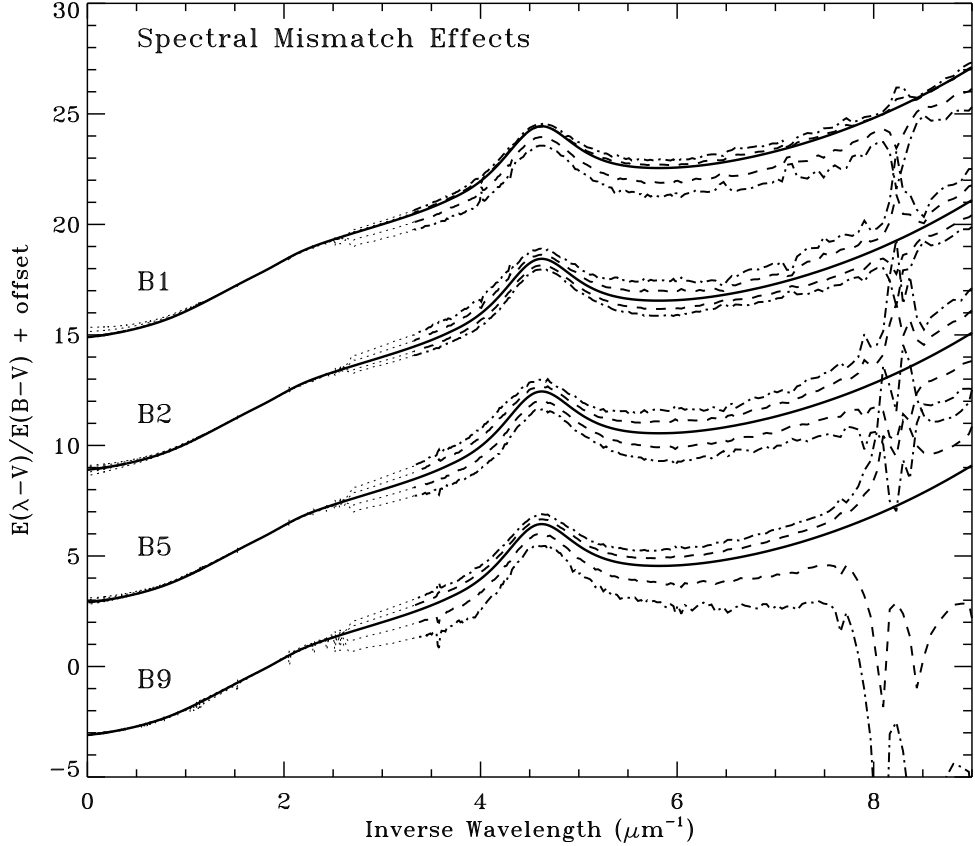}
\caption{Effects of a $\pm \frac{1}{2}$-class spectral mismatch error
on extinction curves derived by the pair method.  In each of the four
groups, the solid curve shows the true shape of a normalized (but
arbitrarily offset) extinction curve affecting a B1, B2, B5, or B9
star.  The dashed and dash-dot curves show the extinction curves
which would be derived via the pair method in the presence of a
$\pm \frac{1}{2}$ spectral type mismatch error in the cases of
$E(B-V)=0.30$ and $E(B-V)=0.15$, respectively.  Detailed
spectrophotometric extinction curves are typically only available in the
UV spectral region, which is highlighted in the figure.  The extension
of the mismatch curves into the optical and IR is shown by the dotted
curves.  Note that the mismatch curves fall below the true curve in the
UV and above the true curve in the IR when the standard star is cooler
than the reddened star, and vice versa.  The feature in the mismatch
curves near $1/\lambda \simeq 2.7 \mu{\rm m}^{-1}$ is due to the Balmer
jump.  The large feature at 8.2 $\mu{\rm m}^{-1}$ (i.e., 1215 \AA) is
due to mismatch of the strong stellar Lyman $\alpha$ absorption line.
In producing these curves, we use the temperature scale: B0.5 =
28000~K, B1 = 25000~K, B2 = 20000~K, B3 = 18000~K, B5 = 15000~K, B7 =
13000~K, B8 = 12000~K, B9 = 11000~K, and A0 = 9500~K.
\label{figMISMATCH}}
\end{figure}


\begin{figure}[ht]
\figurenum{2}
\epsscale{0.85}
\plotone{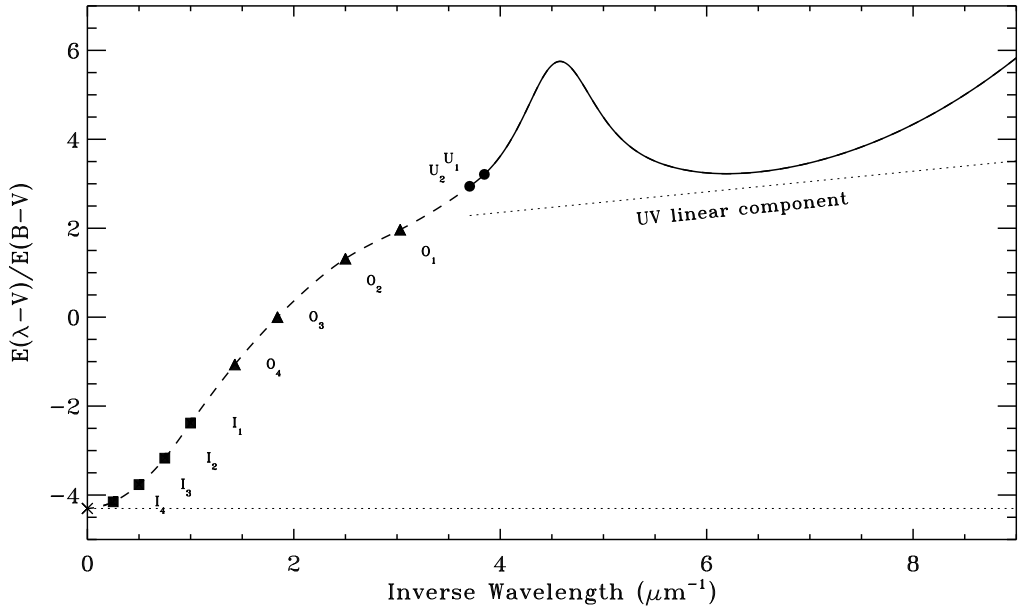}
\caption{A flexible, parametrized representation of the IR through UV
extinction curve.  The curve consists of two parts:  1) the UV (i.e.,
$\lambda \leq 2700$ \AA, solid curve in the figure) where we adopt the
3-component parametrization scheme of FM90; and 2) the optical/IR
(i.e., $\lambda > 2700$ \AA, dashed curve) where we adopt a cubic
spline interpolation through sets of IR ($I_1$,$I_2$, $I_3$, $I_4$),
optical ($O_1$, $O_2$, $O_3$, $O_4$), and UV ($U_1$, $U_2$) ``anchor
points.''  The values of the anchor points and the six parameters
describing the UV portion of the curve are determined by fitting the
observed SED of a reddened star, as described in \S
\ref{secEXTINCTION}.  The dotted line shows the ``linear component'' of
the UV extinction curve.  The other two components, which are added to
the linear component, are the 2175~\AA\ bump profile centered at
$\lambda^{-1} \simeq 4.6
\mu{\rm m}^{-1}$ and the far-UV rise, which refers to the departure of
the curve from the extrapolation of the linear component at very short
wavelengths.  See the discussion in \S \ref{secFLEXI}.  The particular
curve shown in this figure corresponds to that derived in \S
\ref{secFRIENDS} for the star HD~147933.
\label{figFLEXI}}
\end{figure}


\begin{figure}[ht]
\figurenum{3}
\epsscale{0.85}
\plotone{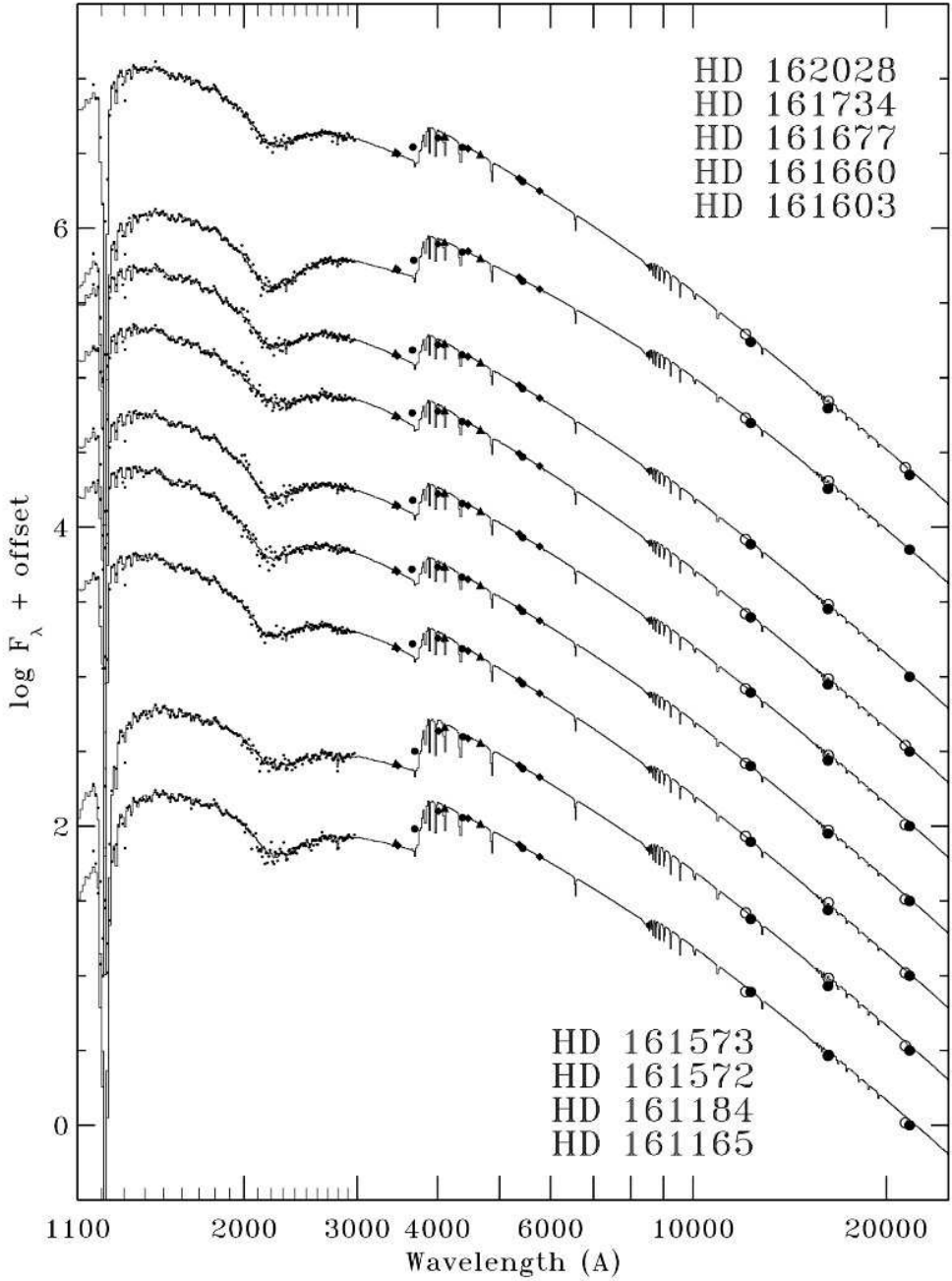}
\caption{SED fits for nine lightly-reddened mid-to-late main sequence B
stars in the open cluster IC 4665.  Histogram-style curves are the
best-fitting, reddened \atlas\/ stellar atmosphere models, arbitrarily
shifted vertically for clarity. In the UV ($\lambda < 3000$ \AA), small
filled circles are \iue\/ low-dispersion spectrophotometry binned to
match the resolution of the models.  In the optical, circles,
triangles, and diamonds are Johnson {\it UBVRI}, Stromgren {\it uvb},
and Geneva {\it UB$_1$B$_2$VV$_1$G} data, respectively. In the IR, the
large open and closed circles are Johnson {\it JHK} and \tmass\/ {\it
JHK} data, respectively.  The photometric data have been converted to
flux units for display purposes only.  As discussed in \S
\ref{secRESULTS}, the comparison between models and photometric
magnitudes, colors, and indices is performed in the native format of
the photometry.
\label{figSED1}}
\end{figure}


\begin{figure}[ht]
\figurenum{4}
\epsscale{0.75}
\plotone{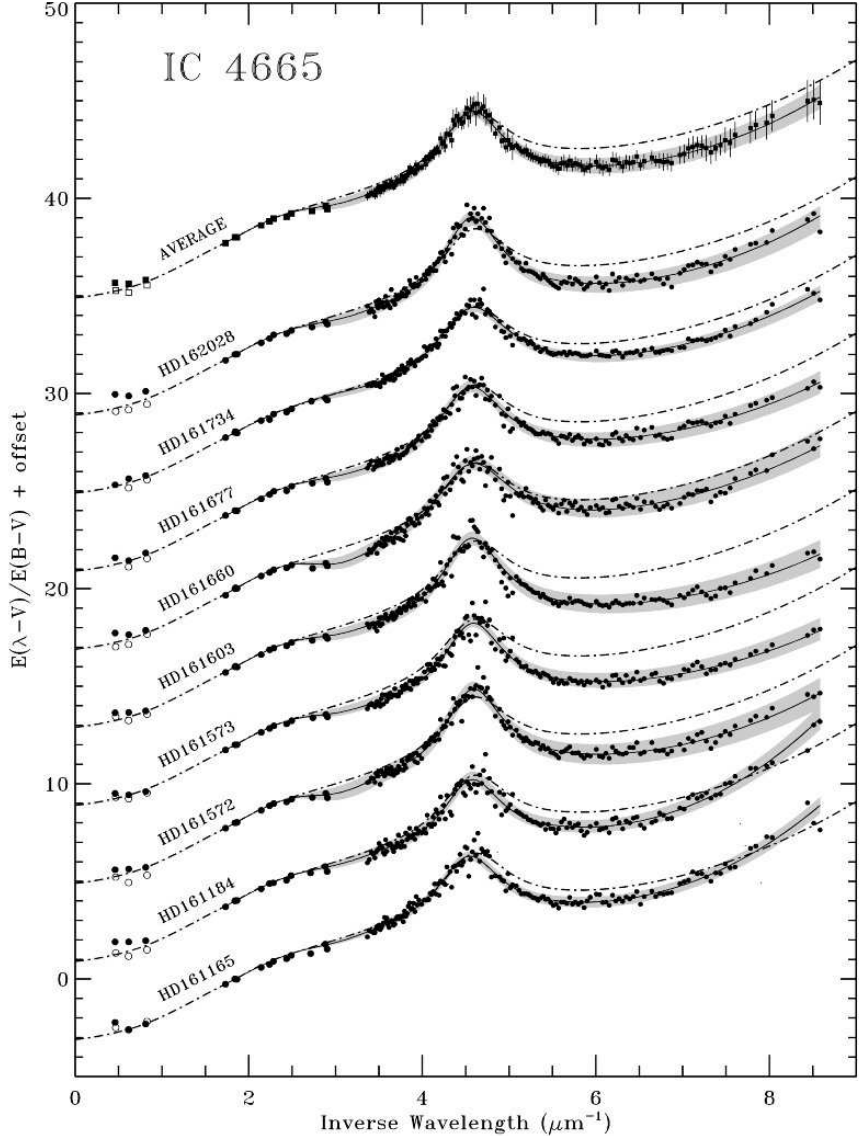}
\caption{\small Extinction curves derived using model atmosphere
calculations for nine
mid-to-late B main sequence stars from the open cluster IC~4665. The
values of $E(B-V)$ range from 0.14 to 0.25 for this group.  The filled
symbols show \iue\/ spectrophotometry in the UV ($\lambda^{-1} > 3.33\/
\mu{\rm m}^{-1}$), Johnson, Stromgren, and Geneva photometry in the
optical, and \tmass\/ {\it JHK} in the near-IR.  Open symbols indicate
near-IR Johnson {\it JHK} data.  The solid curves are the fits to the
data (i.e., the ``flexible, parametrized curves'' described in \S
\ref{secFLEXI}) as determined by the SED-fitting procedure, and the
shaded regions show the 1-$\sigma$ uncertainty in the curves, based on
Monte Carlo simulations.  For comparison, the dash-dot curves show the
average
Galactic extinction curve (corresponding to $R(V) = 3.1$).  Because of
an apparent conflict in the \tmass\/ and Johnson {\it JHK} data, we
assumed that the shape of the IC~4665 curves in the near-IR follows the
average Galactic curve.  The top solid curve and the square symbols
show the simple mean of the nine IC~4665 extinction curves, along with
the average Galactic curve for comparison.  The error bars indicate the
sample standard deviation of the individual curves.  Note that this
scatter is comparable to the 1-$\sigma$ errors of the individual
curves, indicating that the shape of the extinction curve in IC~4665 is
uniform to within our (small) measurement errors.
\label{figEXT1}}
\end{figure}


\begin{figure}[ht]
\figurenum{5}
\plotone{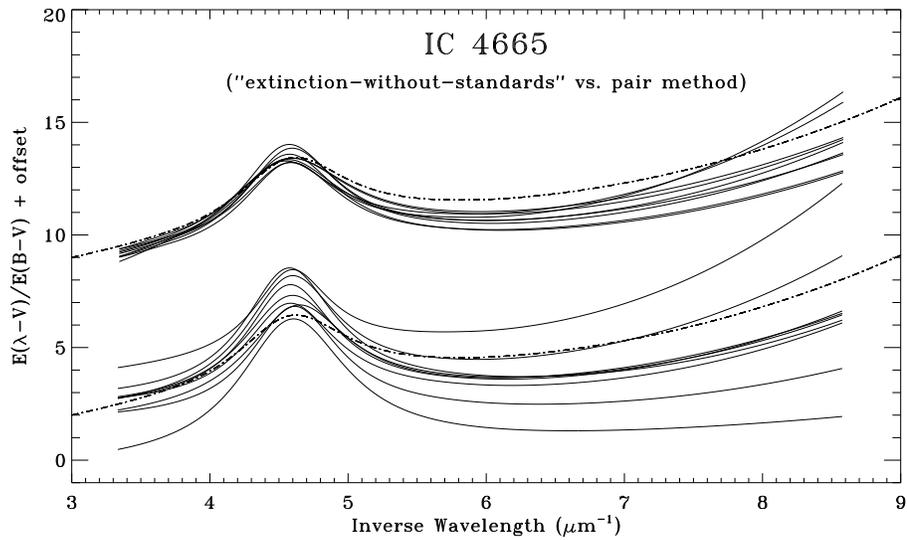}
\caption{Comparison between extinction curves derived for the cluster
IC~4665 using the extinction-without-standards technique described in
this paper (upper set of curves) and using the traditional pair method
approach (lower set of curves; from HHT).  The average Galactic curve
is shown with both sets of curves for reference (dash-dot curves).
\label{figEXT1comp}}
\end{figure}


\begin{figure}[ht]
\figurenum{6}
\epsscale{0.85}
\plotone{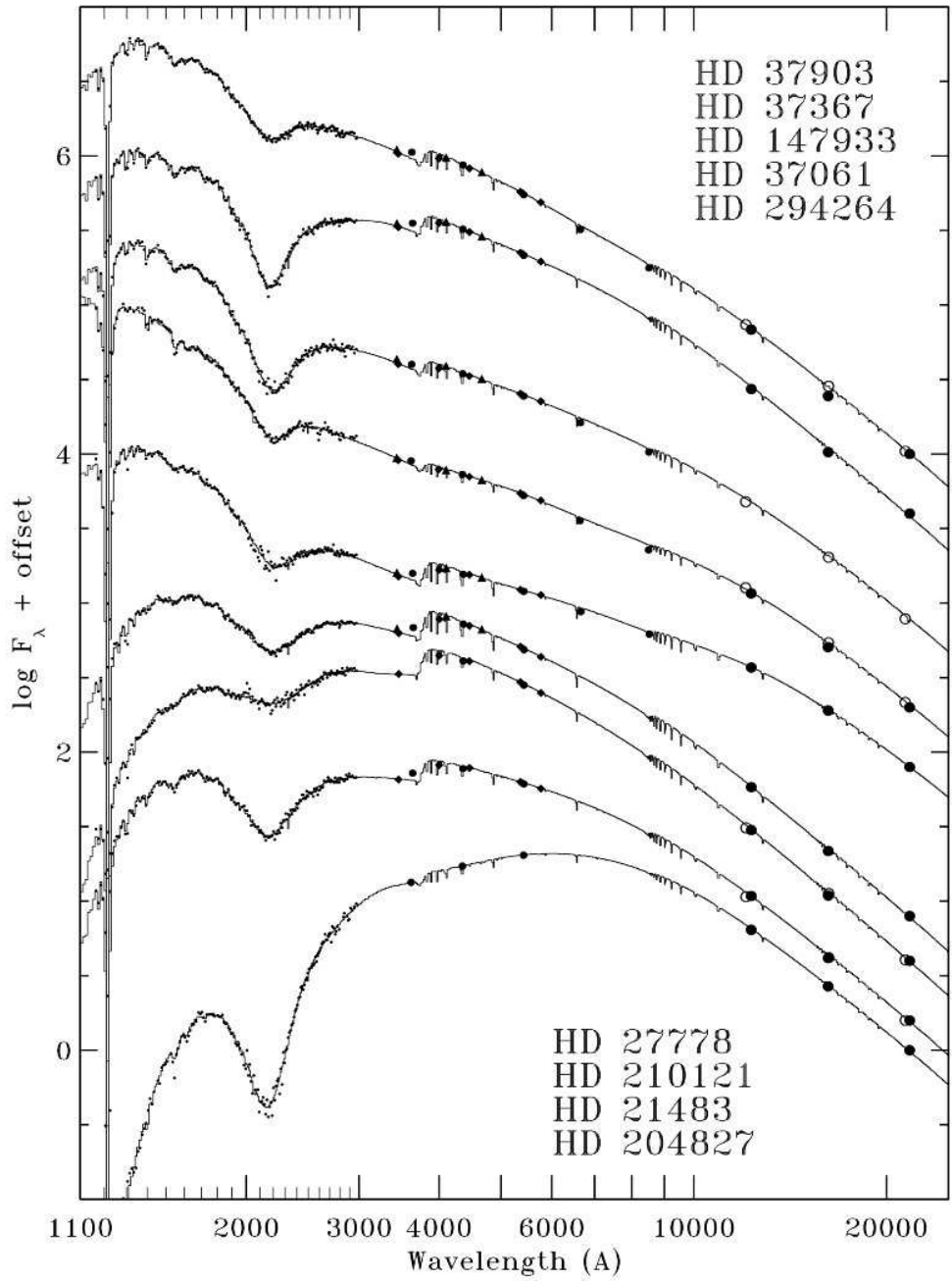}
\caption{Same as Figure \ref{figSED1}, but for a group of nine
moderately
reddened early B stars.
\label{figSED2}}
\end{figure}


\begin{figure}[ht]
\figurenum{7}
\epsscale{0.75}
\plotone{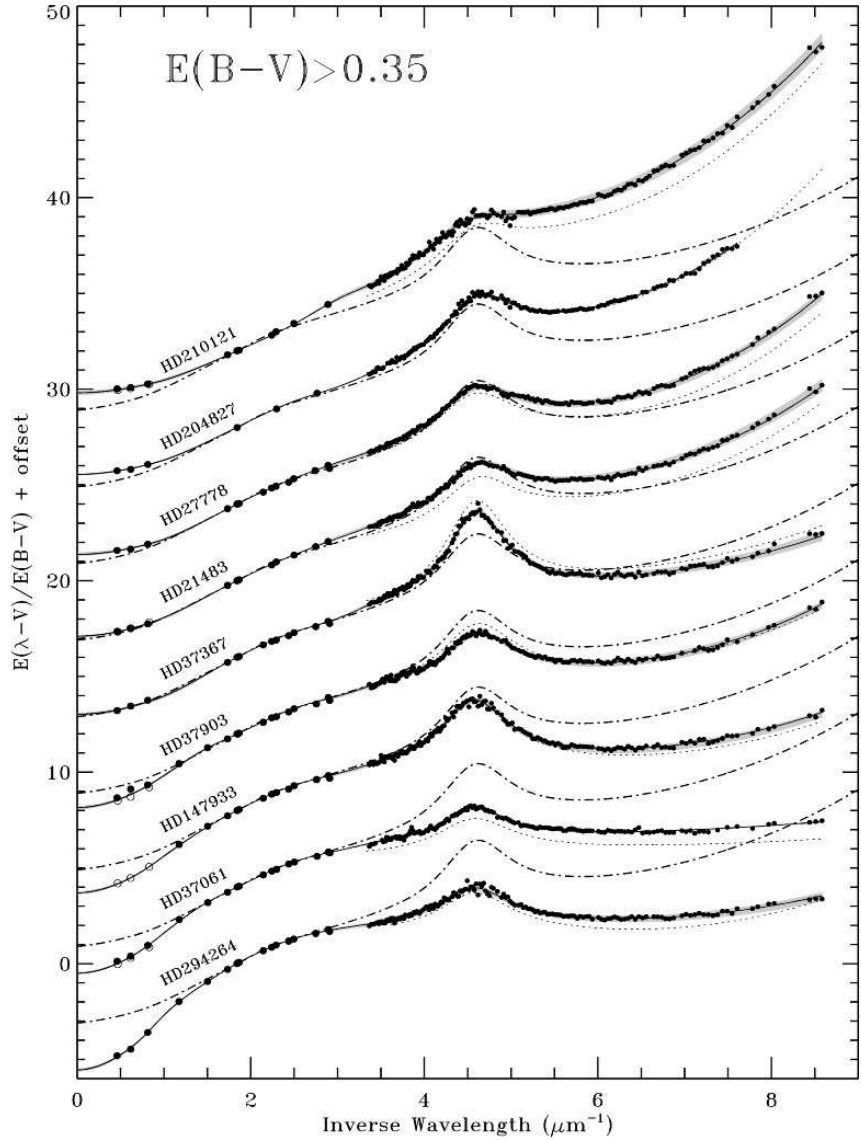}
\caption{Same as Figure \ref{figEXT1}, but for nine moderately reddened
early B stars.  Dotted curves show pair method extinction curves for
all
stars except HD~294264.  Those for HD~210121 and HD~27778 are
unpublished
results by us, and the rest are from the catalog of FM90. We are not
aware of any pair method extinction curves for HD~294264.  The values of
$E(B-V)$ range from 0.36 to 1.08 for this group. These curves
illustrate
the wide range observed in the UV-through-IR wavelength dependence of
interstellar extinction.
\label{figEXT2}}
\end{figure}


\begin{figure}[ht]
\figurenum{8}
\epsscale{0.85}
\plotone{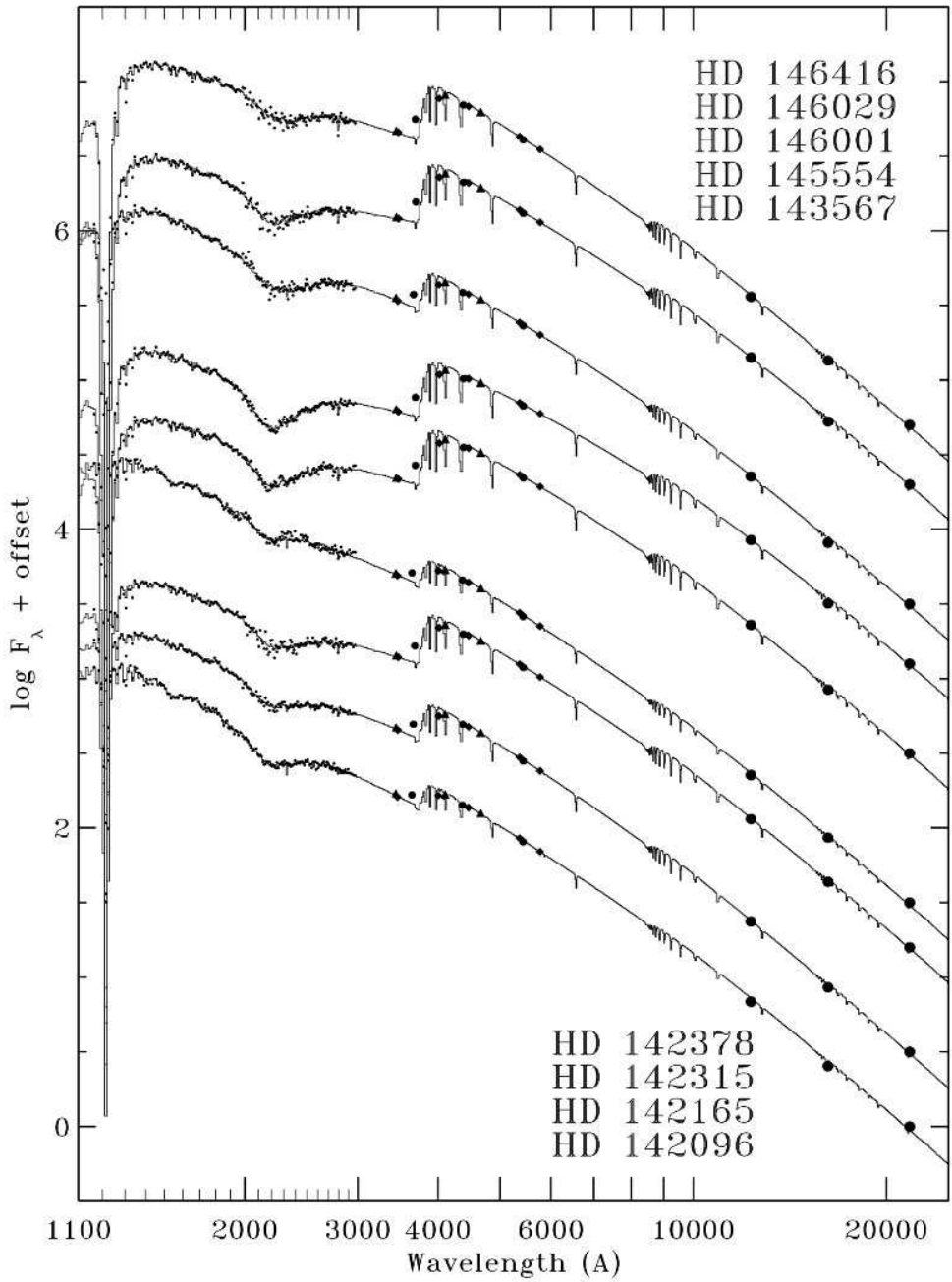}
\caption{Same as Figure \ref{figSED1}, but for a group of nine
lightly-reddened mid-to-late B stars in the direction $347\degr < l <
355\degr$ and $18\degr < b < 26\degr$.
\label{figSED3}}
\end{figure}


\begin{figure}[ht]
\figurenum{9}
\epsscale{0.75}
\plotone{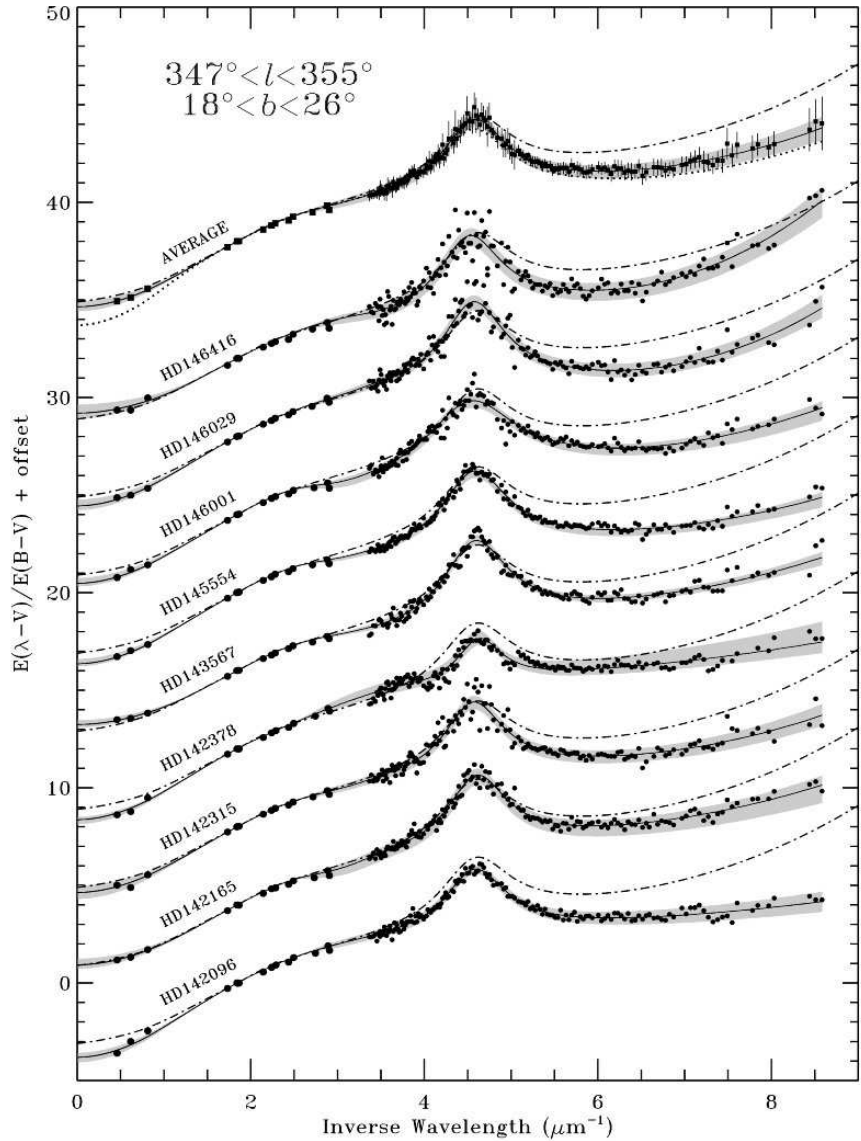}
\caption{Same as Figure \ref{figEXT1}, but for a group of nine
lightly-reddened mid-to-late B stars in the direction $347\degr < l <
355\degr$ and $18\degr < b < 26\degr$.  The values of $E(B-V)$ range
from 0.09 to 0.22 for this group.  The dotted curve, superimposed on
the average curve at the top of the figure  shows the result for the
nearby star HD~147933, located at ($l$, $b$) = ($353.7\degr,17.7\degr$)
with $E(B-V) = 0.50$ (see Figure \ref{figEXT2}).
\label{figEXT3}}
\end{figure}
}
\end{document}